%% file: main_combined.tex
\documentclass[aip,jcp, reprint,preprintnumbers,amsmath,amssymb,floatfix,nofootinbib]{revtex4-1}
\usepackage{graphicx}
\usepackage{color}
\usepackage{amsmath}
\usepackage{mathtools}
\usepackage{amssymb}
\usepackage{amsfonts}
\usepackage{rotating}
\usepackage{multirow}
\usepackage{dcolumn}
\usepackage{bbold}
\usepackage{float}
\usepackage{xr}
\usepackage{dsfont}

\begin{document}

\title{A multi-fragment real-time extension of projected density matrix embedding theory: Non-equilibrium electron dynamics in extended  systems}

\author{D. Yehorova}
 \email{dyehorova3@gatech.edu}

\author{Joshua S. Kretchmer}
 \email{jkretchmer@gatech.edu}

\affiliation{Department of Chemistry and
Biochemistry, Georgia Institute of Technology}

\date{\today}

\begin{abstract}
 
In this work, we derive a multi-fragment real-time extension of projected density matrix embedding theory (pDMET) designed to treat non-equilibrium electron dynamics in strongly correlated systems.  As in the previously developed static pDMET, real time pDMET partitions the total system into many fragments; the coupling between each fragment and the rest of the system is treated through a compact representation of the environment in terms of a quantum bath. Real-time pDMET involves simultaneously propagating the wavefunctions for each separate fragment-bath embedding system along with an auxiliary mean-field wavefunction of the total system. The equations of motion are derived by \emph{(i)} projecting the time-dependent Schr\"{o}dinger equation in the fragment and bath space associated with each separate fragment and by \emph{(ii)} enforcing the pDMET matching conditions between the global 1-particle reduced density matrix (1-RDM) obtained from the fragment calculations and the mean-field 1-RDM at all points in time.
The accuracy of the method is benchmarked through comparisons to time-dependent density-matrix renormalization group (TD-DMRG) and time-dependent Hartree-Fock (TDHF) theory; the methods were applied to a one- and two-dimensional single-impurity Anderson model and multi-impurity Anderson models with ordered and disordered distributions of the impurities.
The results demonstrate a large improvement over TDHF and rapid convergence to the exact dynamics with an increase in fragment size. 
Our results demonstrate that real-time pDMET is a promising and flexible method that balances accuracy and efficiency to simulate non-equilibrium electron dynamics in heterogeneous systems of large size.

\end{abstract}


\maketitle

\section{Introduction}

Recent developments in time-resolved spectroscopy have dramatically expanded the opportunity for real-time observation of correlated electron dynamics in complex molecular environments at the attosecond scale\cite{Pen19, nis17, Var22, Gal12}. Phenomena of interest include charge transfer following photoexcitation in biological\cite{Maa21, De19} and photovoltaic systems\cite{Kru20, Fue16, ponseca_ultrafast_2017}, electron and spin transport in molecular junctions\cite{cui_thermal_2019, joachim_electronics_2000, nitzan_electron_2003, zhang_recent_2021, dayen_two-dimensional_2020} or quantum dots \cite{singh_hot-electron_2019, bande_electron_2013}, Auger processes in small molecules and surfaces,\cite{jahnke_interatomic_2020, demekhin_interatomic_2009, ouchi_interatomic_2011}, among others. 
New experimental techniques and a wide range of applications make development of real-time non-equilibrium electronic structure methods relevant, while bringing the challenge of accurate treatment of electron correlation and large system size to the forefront.

Some of the methods that are currently used to address this question include techniques based on non-equilibrium Green's functions\cite{Aok14, Har08, Ram18, Fre06}, semi-classical approximations\cite{Li13, Swe11, Swe12}, real-time Monte Carlo methods\cite{Coh14a, Coh14b, Gul11, Sch10a, Wer09,Coh15}, numerical path-integral techniques\cite{Muh08, Wei09, Seg10b} and wave function propagation-based methods\cite{li_real-time_2020, goings_real-time_2018, schlunzen_ultrafast_2019, church_real-time_2021, thiel_semiempirical_2014, chatterjee_real-time_2019}. While methods based on frequency domain response theory showed great success in the calculation of response properties of complex systems\cite{helgaker_recent_2012, norman_simulating_2018, vtoukach_recent_2013}, real-time approaches show a great advantage in the description of nonlinear phenomena and provide versatility in the calculated observables\cite{goings_real-time_2018}.

Methods like time dependent density matrix renormalization group (TD-DMRG)\cite{Bai19, Fra19, Bai21, li_finite-temperature_2020}, configuration interaction (TD-CI)\cite{sonk_td-ci_2011} and other complete active space based time-dependent methods \cite{Dur22, Pen18, Sat13, mir11, kochetov_rhodyn_2022} achieve high accuracy and have been highly applicable in the spectroscopic studies of small molecular systems. However, their practical application is limited by high computational cost. Real-time time-dependent density functional theory can be a powerful tool that decreases computational cost associated with the large-scale systems, but the inherent difficulty of reproducing some of the fundamental effects associated with approximations of the exchange-correlation potential leaves room for further improvements in real-time methods\cite{li_real-time_2020, isborn_time-dependent_2007-1}.

In this work we address the challenge associated with treatment of a large system at a high level of accuracy through the use of quantum embedding. Embedding techniques are based on the highly accurate treatment of a small region of interest (subsystem) and an approximate, inexpensive calculation of the surrounding environment. In particular, we extend a specific form of density matrix embedding theory (DMET)\cite{Kni12, Kni13, Wou16}, projected DMET (pDMET)\cite{wu_projected_2019}, for the treatment of non-equilibrium electron dynamics. 

Both forms of DMET are based on forming a compact, but approximate description of the environment in terms of a set of embedding orbitals derived from the Schmidt decomposition of a mean-field wavefunction for the total system. This can be performed either for a single subsystem or in a multi-fragment formulation that involves multiple subsystems each coupled to their own set of embedding orbitals.\cite{Wou16,wu_projected_2019} The difference between conventional DMET and pDMET is the treatment of self-consistency between the mean-field calculation and the correlated calculations of the subsystem and embedding orbitals. In the original formulation of DMET, this self-consistency is achieved by adding the correlation potential, a single-particle operator, to the Fock operator of the total system; the terms of the correlation potential are varied to minimize the difference between the one-particle reduced density matrices (1-RDM) of the mean-field and correlated calculations.\cite{Wou16} In comparison, the self-consistency condition is achieved in pDMET by projecting a global 1-RDM, constructed by averaging over the correlated 1-RDMs of all fragments, onto the set of idempotent matrices. While the original DMET matching condition provided a numerical relationship between the mean-field and correlated calculations, pDMET introduces an analytical correspondence between the global and mean-field 1-RDMs. DMET has been successfully applied to a variety of strongly correlated model systems\cite{zheng_ground-state_2016, cui_efficient_2020, faulstich_pure_2022, reinhard_density-matrix_2019}, extended materials  \cite{pham_periodic_2020, cui_efficient_2020} and spin systems\cite{fan_cluster_2015, gunst_block_2017}. It has also been extended to excited state calculations \cite{Tra19, mitra_excited_2021, qiao_density_2021}, bond breaking events\cite{pham_can_2018}, and electron-phonon\cite{sandhoefer_density_2016} and long range interactions\cite{ricke_performance_2017}. 

 In the context of dynamics, a real-time expression of DMET has been previously formulated and was shown to have excellent results in the context of the single impurity Anderson model\cite{kretchmer_real-time_2018}. The equations of motion for the embedding wave-function were derived using the time-dependent variational principle (TDVP). However, the variational nature of the equations of motion does not include a time-dependent condition that ensures consistency between fragments, which limits the method to a single-fragment formulation. 
 
 In principle, a correlation potential can be introduced in a time-dependent fashion to enforce self-consistency between the mean-field and correlated 1-RDMs throughout time. However, it was found that $v$-representability issues almost always occurred after sufficiently long time-propagation, such that the dynamics became completely independent of the correlation potential\cite{kretchmer_real-time_2018}. 
 
 Therefore, in this work, we take advantage of the analytic matching condition in pDMET to derive a real-time formulation of pDMET which eliminates the need for the TDVP and allows for a multi-fragment implementation. This is achieved by simultaneously propagating the correlated wavefunctions associated with each fragment and an auxiliary dynamics corresponding to a mean-field dynamics for the total system, while simultaneously enforcing the pDMET matching conditions at each point in time.  The derived equations of motion provide a unique real-time method for the simulation of non-equilibrium electron dynamics that balances accuracy and computational efficiency in extended systems. Each fragment is small in size, such that the the accurate dynamics are only solved in a compact space. Furthermore, the dynamics within each fragment can be treated simultaneously such that the method is parallelizable to allow for efficient scaling with total system size.

 To benchmark the real-time pDMET method, we apply it to the single- and multi-impurity Anderson model in one and two-dimensions, where the impurities are distributed both in ordered and disordered fashions. We observe a marked improvement over time-dependent Hartree-Fock (TDHF) theory and excellent agreement with the numerically exact TD-DMRG, while only explicitly correlating a small fraction of the total number of orbitals. This agreement occurs even in the strongly correlated regime for disordered systems, for which a single-impurity formulation of the method is poorly suited.

\section{Method}

In this section, we begin by first summarizing the static pDMET equations necessary to provide a foundation for the subsequent presentation of the real-time formulation; we direct readers to previous work for the full details of the static pDMET method.\cite{wu_projected_2019} The following notation will be utilized throughout all sections: the site basis, which is comprised of the orthonormal real-space localized single-particle basis, will be indexed as $p, q, r, s$; the embedding basis, which is comprised of the correlated set of subsystem and associated embedding orbitals, are indexed by $a, b, c$ and $d$; and the natural orbital basis, which is comprised of the eigenvectors of the correlated global 1-RDM formed from the 1-RDMs of each fragment, are indexed by $\mu$ and $\nu$.

\subsection{Static pDMET}

Let us consider a system spanned by an orthonormal single-particle basis, where $N$ is the number of basis functions and $N_e$ is the total number of electrons in the system. These basis function will be further referred to as \emph{sites}. The general Hamiltonian of the system is:
\begin{equation}
    \hat{H}=\sum_{pq}^{N}\sum_{\sigma}h_{pq}\hat{a}^\dag_{p\sigma}\hat{a}_{q\sigma}+\frac{1}{2}\sum_{pqrs}V_{pqrs}\sum_{\sigma\tau}\hat{a}^\dag_{p\sigma}\hat{a}_{r\tau}^{\dag}\hat{a}_{s\tau}\hat{a}_{q\sigma},\label{eqn:ham_gen}
\end{equation}
where $h_{pq}$ and $V_{pqrs}$ are one- and two - electron integrals and $\hat{a}^\dag_{p\sigma}$ and $\hat{a}_{p\sigma}$ are creation and annihilation operators associated with site $p$ and spin $\sigma$, respectively. 

\begin{figure}
\centering
\includegraphics[scale=0.50]{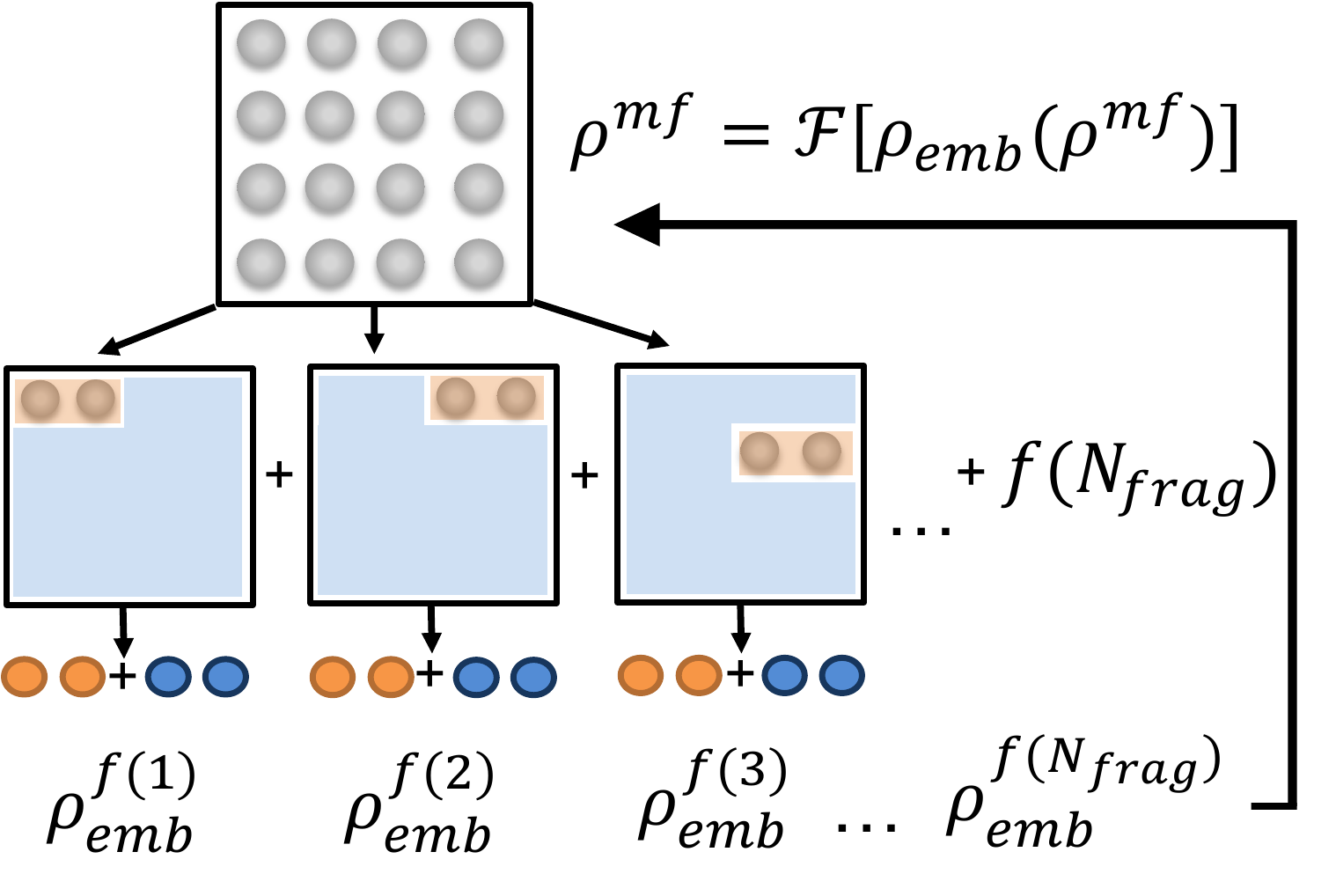}
\caption{Schematic representation of multi-fragmented static DMET. The system is divided into $N_{frag}$ fragments, each containing $N_{sub}$ subsystem orbitals (pictured in orange). The environment of each fragment is described in terms of $N_{sub}$ bath orbitals (pictured in blue) that are correlated with the subsystem orbitals. These bath orbitals are obtained from an idempotent density matrix, $\rho^{mf}$, of the total system. An independent correlated calculation is performed in the subsystem-bath space of each fragment to form a correlated density matrix, $\rho_{emb}$. The mapping $\mathcal{F}$ takes in the set of correlated $\rho_{emb}$ and generates a new idempotent $\rho^{mf}$. The form of the mapping $\mathcal{F}$ is the core difference between pDMET and previous formulations of the method.}
\label{fig:fragments}
\end{figure}

In the multi-fragment formulations of DMET or pDMET, the total system is partitioned into $N_{frag}$ fragments that each consist of $N_{sub}$ subsystem orbitals. Each fragment is treated as an independent embedding problem where the orbitals that span the fragment are defined as a subsystem and the rest of the system is its environment, pictorially illustrated in Figure \ref{fig:fragments}. For each fragment, the global wavefunction can be compactly written through its Schmidt decomposition, such that
\begin{equation}
    |\Psi\rangle= \sum_{\alpha=1}^{N_A}\lambda_{\alpha}|A_{\alpha}\rangle|B_{\alpha}\rangle
    \label{eq:Sch_decomposition}
\end{equation}
where $|A_{\alpha}\rangle$ are multi-electron states describing the subsystem A and $|B_{\alpha}\rangle$ are a special set of multi-electron states describing the environment. Note that the number of states used to describe the environment is given by $N_A$: the same number of multi-electron states used to describe the subsystem. If the states $|B_{\alpha}\rangle$ are obtained from the Schmidt decomposition of the exact global wavefunction, they exactly capture the entanglement with the environment.

 In practical DMET calculations, the states $|B_\alpha\rangle$ are obtained from the Schmidt decomposition of a mean-field wavefunction for the total system, $|\Psi^{mf}\rangle$. The global wavefunction is thus given by a single Slater determinant and can be described in terms of its single-particle density matrix,
\begin{equation}
   {\rho}^{mf}_{pq}=\langle\Psi^{mf}|\sum_\sigma\hat{a}^\dag_{q\sigma}\hat{a}_{p\sigma}|\Psi^{mf}\rangle.
   \label{eq:rho_mf_def}
\end{equation}
%
In this case, the states $|B_\alpha\rangle$ are comprised of determinants formed from a set of \emph{single-particle} embedding orbitals. These embedding orbitals can be obtained in many mathematically equivalent ways.\cite{wouters_practical_2016} Here, we obtain the embedding orbitals associated with fragment $A$ by diagonalizing the portion of $\rho^{mf}$ that corresponds solely to the environment of fragment $A$, such that
\begin{equation}
    \sum_{q \in env}^{N-N_{sub}}\rho^{mf}_{pq}(R^{emb}_{qa})^A = \lambda^{A}_{a}(R^{emb}_{pa})^A,
    \label{eq:R_and_rho_mf}
\end{equation}
where both $p$ and $q$ are restricted to the environment of fragment $A$; $p$ and $q$ run over all indices except those that correspond to the subsystem orbitals of fragment $A$. The $N-N_{sub}\times N-N_{sub}$ matrix $(R^{emb})^A$ corresponds to the matrix of expansion coefficients that define the embedding orbitals in terms of a linear combination of the site orbitals associated with the environment of fragment $A$.

The embedding orbitals, $(R^{emb})^A$, can be classified into three types based on their corresponding eigenvalues  in Eq. \eqref{eq:R_and_rho_mf}. Bath orbitals have eigenvalues between zero and two and are fully correlated with the subsystem orbitals. Core orbitals have an eigenvalue equal to two and are fully occupied in the states $|B_\alpha\rangle$. And virtual orbitals have an eigvenvalue equal to zero and are fully unoccupied in the states $|B_\alpha\rangle$. This implies that the full embedding wavefunction associated with each fragment is a complete active space (CAS)-like wavefunction composed of the subsystem and corresponding embedding orbitals.\cite{wouters_practical_2016}

In this work we use the following notation for the CAS-like embedding wavefunction associated with fragment $A$:
\begin{equation}
    |\Psi^A\rangle=\sum_{M}C^A_{M}|M^A\rangle
    \label{eq:frag_CAS_wfcn},
\end{equation}
where $|M\rangle$ denotes a determinant in the embedding space and $C_M$ the corresponding expansion coefficient.

In the static algorithms, the CAS-like embedding wavefunction for each fragment can be calculated by finding the ground-state of an embedding Hamiltonian, $H_{emb}^A$.\cite{wouters_practical_2016} The embedding Hamiltonian is obtained by projecting the original fully interacting Hamiltonian, Eq. \eqref{eqn:ham_gen}, into the CAS space. This projection can be performed by performing a change of single-particle basis using the following rotation matrix, while including a contribution from the core orbitals,\cite{wouters_practical_2016}

\begin{gather}
 R^{A}
 =
  \begin{bmatrix}
   \mathds{1}_{N_{sub} \times N_{sub}} & 0 \\
   0 & (R^{emb})^A
   \end{bmatrix}
\end{gather}
Here, the identity matrix denotes that the sub-system orbitals are the same in the original site basis and the embedding basis; $(R^{emb})^A$ is  the matrix of embedding orbitals and is defined in Eq. \eqref{eq:R_and_rho_mf}.
 
The final step in the DMET algorithm is introducing a self-consistency between the correlated fragment calculations and the original mean-field calculation used to generate the embedding orbitals. For the specific case of pDMET this is accomplished by first constructing a correlated 1-RDM in the global domain, $\rho^{glob}$, using the 1-RDMs from all separate fragment calculations.\cite{wouters_practical_2016,wu_projected_2019}

\begin{equation}
\begin{aligned}
    {\rho}^{glob}_{pq}={} & \sum_{ab}\frac{1}{2}\left[R_{pa}^{f(p)}{(\rho}_{ab}^{emb})^{f(p)}R^{*f(p)}_{qb}\right.\\
                      & + \left.R_{pa}^{f(q)}({\rho}^{emb}_{ab})^{f(q)}R^{*f(q)}_{qb}\right]
    \label{eq:_rho_glob_def}
\end{aligned}
\end{equation}
The index $f(p)$ corresponds to the fragment that includes site $p$ and $({\rho}^{emb})^{f(p)}$ corresponds to the 1-RDM obtained from the embedding wavefunction for fragment $f(p)$,
\begin{eqnarray}
({\rho}^{emb}_{ab})^{f(p)}&=&\langle\Psi^{f(p)}|\sum_\sigma\hat{a}^\dag_{b\sigma}\hat{a}_{a\sigma}|\Psi^{f(p)}\rangle.\label{eq:rho_emb_def}
\end{eqnarray}

The global 1-RDM, Eq. \eqref{eq:_rho_glob_def}, is projected onto the space of idempotent matrices with $N_e$ electrons to form a new guess for the mean-field wavefunction of the total system.\cite{wu_projected_2019} In the spin restricted formulation, which is utilized in this work, the corresponding idempotent matrix is formed from the $N_{occ}=N_e/2$ eigenvectors of $\rho^{glob}$ with the highest occupation. 
These eigenvectors of $\rho^{glob}$ are referred to as natural orbitals, represented by the matrix $U$, and indexed by $\mu$ and $\nu$, such that
\begin{equation}
    \sum_{q}^{N}\rho^{glob}_{pq}U_{q\mu} = \lambda_{\mu}U_{q\mu}
    \label{eq:N_and_rho_glob}
\end{equation}\
and
\begin{equation}
    \rho^{mf}_{pq} = 2 \sum_{\mu}^{N_{occ}}U_{p\mu}U_{q\mu}^{*}
    \label{eq:rho_mf_and_rho_glob}
\end{equation}
Given the newly formed $\rho^{mf}$, the pDMET algorithm is iterated until $\rho^{mf}$ does not change between iterations.

The alternative self-consistency procedure used in pDMET compared to conventional DMET results in a significant reduction of the computational cost for a large non-periodic lattice and provides a more robust analytical relationship between the embedding and mean field representations of the problem. The latter point is crucial for the derivation of a real-time multi-fragmented DMET algorithm. 

\subsection{Real-time pDMET}

We now describe the central methodological contribution of the work, a multi-fragment real-time extension of pDMET. This is achieved through the simultaneous propagation of \emph{(i)} all the CAS-like wavefunctions for each fragment along with \emph{(ii)} an additional mean-field dynamics for the total system. The equations of motions are derived to maintain the inherit relationship between the mean-field and global description of the system. This is achieved by preserving the pDMET self-consistencies that connect the mean-field and embedding representations of the system at each point in time. 
The full derivations of the working dynamical equations defined in this section can be found in the Supplemental Information (SI).

We begin by first deriving the formal time-dependence of the embedding wavefunction of fragment $A$ by projecting the time-dependent Schr\"{o}dinger equation into the embedding space,
\begin{eqnarray}
i|\dot{\Psi}^A\rangle&=&\hat{H}|\Psi^A\rangle\\
iP^A|\dot{\Psi}^A\rangle&=&P^A\hat{H}P^A|\Psi^A\rangle\\
i\frac{\partial}{\partial t}\left(P^A|\Psi^A\rangle\right)-i\dot{P}^A|\Psi\rangle&=&\hat{H}^{emb}|\Psi^A\rangle\\
i|\dot{\Psi}^A\rangle=\hat{H}^{emb}|\Psi^A\rangle&+&i\dot{P}^A|\Psi^A\rangle\label{eq:formal_time_dep_of_wfcn}
\end{eqnarray}
where $P^A$ is the formal projector into the embedding space of fragment $A$, $H^{emb}=P^AHP^A$ is the Hamiltonian projected into the embedding space, and we have used that $P^A|\Psi^A\rangle=|\Psi^A\rangle$\cite{wouters_practical_2016}. Eq. \eqref{eq:formal_time_dep_of_wfcn} indicates that the time-dependence of the embedding wavefunction has two components, the first associated with the embedding Hamiltonian and the second associated with the time-dependence of the quantum bath given by $\dot{P}^A$.

For the case of a mean-field global wavefunction, $|\Psi^A\rangle$ is given by Eq. \eqref{eq:frag_CAS_wfcn}, and the projector is $P=\sum_M|M^A\rangle\langle M^A|$. Therefore, the time-dependence of $|\Psi^A\rangle$ is described in terms of the time-dependence of the expansion coefficients, $C^A$, and determinants, $|M^A\rangle$,
\begin{eqnarray}
i|\dot{\Psi}^A\rangle &=& \sum_Mi\dot{C}^A_M|M^A\rangle+C^A_Mi|\dot{M}^A\rangle\\
&=&\sum_Mi\dot{C}^A_M|M^A\rangle+C^A_M\hat{X}^A|M^A\rangle,\label{eq:time_dep_of_wfcn}
\end{eqnarray}
and the time-dependence of the projector is given by
\begin{eqnarray}
i\dot{P^A}&=&\hat{X}^AP^A-P^A\hat{X}^A.\label{eq:time_dep_of_P}
\end{eqnarray}
Here, we have introduced the single-particle operator $\hat{X}$ that governs the time-dependence of the embedding orbitals that comprise the determinants, such that
\begin{equation}\label{eq:td_R}
    i\dot{R}^A_{pa}=\sum_{q}X^A_{pq}R^A_{qa}.
\end{equation}
This operator is analogous to the operator employed in time-dependent CASSCF (TD-CASSCF) that propagates the orbitals of the CASSCF-wavefunction.\cite{sato_time-dependent_2013}

Equation \eqref{eq:time_dep_of_wfcn} indicates that the dynamics of the embedding wavefunctions are given by the equations of motion for the expansion coefficients $C^A$ and the matrix elements of the operator $\hat{X}^A$. The equations of motion of the coefficients $C^A$ are derived by combining Eqs. \eqref{eq:formal_time_dep_of_wfcn}, \eqref{eq:time_dep_of_wfcn}, \eqref{eq:time_dep_of_P} and multiplying by $\langle N^A|$ yielding
\begin{equation}\label{eq:time_dep_of_C}
    i\dot{C}^A_{N}=\sum_{M}(H^{emb}_{NM}-X^A_{NM})C^A_{M}.
\end{equation}
 This equation effectively corresponds to the time-dependent full CI equations of motion in the compact embedding space. In principle, further computational efficiency could be obtained by restricting the number of allowed excitations in $\hat{H}^{emb}$ and $\hat{X}^A$, albeit at decreased accuracy.

We now turn our attention to the derivation of the matrix elements of the operator $\hat{X}^A$. First, in order to maintain the original definition of the fragments $A$, the subsystem orbitals are defined as time-independent. Therefore, $X^A_{pq}=0$ if orbitals $p \textrm{ or } q \in A$. Non-subsystem matrix elements of the operator $\hat{X}^A$ are derived by taking the time-derivative of the pDMET condition in Eq. \eqref{eq:R_and_rho_mf}, which yields
\begin{equation}\label{eq:X_matrix}
X_{ba}^A=\sum_{pq}\frac{R_{pb}^{A*}i\dot{\rho}^{mf}_{pq}R_{qa}^A}{\lambda_{a}^A-\lambda_{b}^A} 
\end{equation}
This matrix can be interpreted as a rotation of all embedding pDMET orbitals in time, which allows for a few simplifications. Analogous to TD-CASSCF, any intraspace rotations among embedding orbitals of the same type are redundant.\cite{sato_time-dependent_2013} 
We can therefore choose the elements of $\hat{X}^A$ for these rotations. We make the common choice to set $X^A_{ab}=0$ if $a$ and $b$ both correspond to either core orbitals or virtual orbitals. However, unlike TD-CASSCF, the bath-bath rotations are not arbitrary, and are given by Eq. \eqref{eq:X_matrix}, because we have the additional constraint that the embedding orbitals must be eigenvectors of the mean-field 1-RDM as given in Eq. \eqref{eq:R_and_rho_mf}; this is not an issue for the core and virtual space as those orbitals are degenerate. Additionally, though, we are able to set $X^A_{ab}=0$ for $a=b$ even for the bath-orbitals as the diagonal elements of $\hat{X}^A$ only affect the norm of the orbitals.

The time dependence of $\rho^{mf}$ referenced in equation \eqref{eq:X_matrix} is derived by enforcing the pDMET matching condition between the natural orbitals of the $\rho^{glob}$ and $\rho^{mf}$, Eq. \eqref{eq:rho_mf_and_rho_glob}. To derive the time-dependence of $\rho^{mf}$ we first introduce a single-particle operator, $\hat{G}$ that governs the time-dependence of the natural orbitals:
\begin{equation}\label{eq:td_U}
     i\dot{U}_{p\mu}=\sum_{q}G_{pq}U_{q\mu}.
\end{equation}
The time-dependence of $\rho^{mf}$ can be obtained by taking the time-derivative of Eq. \eqref{eq:rho_mf_and_rho_glob} and using Eq. \eqref{eq:td_U}, yielding
\begin{equation}\label{eq:td_rho_mf}
    i\dot{\rho}^{mf}_{pq}=\sum_{r}[G_{pr}\rho^{mf}_{rq}-\rho^{mf}_{pr}G_{rp}].
\end{equation}

The matrix-elements of $\hat{G}$ are derived from the relationship between the global 1-RDM and the natural orbitals, Eq. \eqref{eq:N_and_rho_glob}. Differentiating Eq. \eqref{eq:N_and_rho_glob} with respect to time and solving for the matrix-elements of $\hat{G}$ yields
\begin{equation} \label{eq:G_def}
    G_{\mu\nu}=\sum_{pq}\frac{U_{p\mu}^{*}i\dot{\rho}^{glob}_{pq}U_{q\nu}}{\lambda_{\nu}-\lambda_{\mu}}
\end{equation}
where $G_{\mu\nu}$ is defined in the natural orbital basis. The matrix elements in the site-basis utilized in Eq. \eqref{eq:td_rho_mf} can be obtained by a single-particle rotation using the natural orbitals, $U$, as the rotation matrix. The matrix elements of $\hat{G}$ take on a similar form as $\hat{X}$ due to the analogous structure of Eqs. \eqref{eq:R_and_rho_mf} and \eqref{eq:N_and_rho_glob}.

Analogous to the definition of $X^A_{ba}$, $G_{\mu\nu}=0$ for $\lambda_\mu=\lambda_\nu$. Additionally, a regularization parameter $\varepsilon_G$ is introduced such that $G_{\mu\nu}=0$ if $|\lambda_\mu-\lambda_\nu|<\varepsilon_G$ to avoid numerical instabilities in the dynamics. Similar parameters have been utilized in the context of TD-CASSCF and MCTDH.\cite{kretchmer_real-time_2018,Kre18b,manthe_wavepacket_1992,meyer_multi-configurational_1990}

The mean-field dynamics retain an implicit dependence on the time-evolution of $\rho^{glob}$ through the form of $\hat{G}$, Eq. \eqref{eq:G_def}. The dynamics of $\rho^{glob}$ are governed by the correlated dynamics of each fragment through differentiation of Eq. \eqref{eq:_rho_glob_def},

\begin{widetext}
\begin{eqnarray}\label{eq:td_rho_glob}
    i\dot{\rho}^{glob}_{pq}&=& \frac{1}{2}\sum_{abc}\left[{R}_{pc}^{f(p)}X_{ca}^{f(p)}(\rho_{ab}^{emb})^{f(p)}R^{*f(p)}_{qb}-{R}_{pa}^{f(p)}(\rho_{ab}^{emb})^{f(p)}X_{bc}^{f(p)}{R}^{*f(p)}_{qc}\right.\nonumber\\
     &&+\left.{R}_{pc}^{f(q)}X_{ca}^{f(q)}(\rho_{ab}^{emb})^{f(q)}R^{*f(q)}_{qb}-{R}_{pa}^{f(q)}(\rho_{ab}^{emb})^{f(q)}X_{bc}^q{R}^{*f(q)}_{qc}\right]\nonumber\\
     &&+\frac{1}{2}\sum_{ab}\left[R_{pa}^{f(p)}i(\dot{\rho}_{ab}^{emb})^{f(p)}R^{*f(p)}_{qb} + R_{pa}^{f(q)}i(\dot{\rho}_{ab}^{emb})^{f(q)}R^{*f(q)}_{qb}\right].
\end{eqnarray}
\end{widetext}
Here we have used Eq. \eqref{eq:td_R}.

The time-dependence of $(\rho^{emb})^A$ is obtained by \emph{(i)} differentiating Eq. \eqref{eq:rho_emb_def} and \emph{(ii)} simplifying the expression based on the form of the 1-RDM and 2-RDM for the CAS-like fragment wavefunction, analogously to the derivation of the TD-CASSCF and static CASSCF theories.\cite{sato_time-dependent_2013, Hel14} The final expression for the time-dependence of $(\rho^{emb})^A$ is given by

\begin{eqnarray}\label{eq:td_rho_corr}
       i(\dot{\rho}^{emb}_{ab})^A&=&
       (F_{ba}-F_{ab}^*)^A\nonumber\\
       &+&\sum_c\left((\rho_{ac}^{emb})^AX_{cb}^A-X_{ac}^A(\rho^{emb}_{cb})^A\right)
\end{eqnarray}
where the matrix $F$ is the generalized Fock matrix.\cite{Hel14}
    
Plugging Eq. \eqref{eq:td_rho_corr} into Eq. \eqref{eq:td_rho_glob} and simplifying yields the final expression for $\dot{\rho}^{glob}$,
\begin{eqnarray}\label{eq:td_rho_glob_final}
i\dot{\rho}^{glob}_{pq}&=& \sum_{ab}\frac{1}{2}\left[R_{pa}^{f(p)}( F_{ba}-F_{ab}^*)^{f(p)}R^{f(p)}_{qb}\right.\nonumber\\
                     &&+ \left.R_{pa}^{f(q)}( F_{ba}-F_{ab}^*)^{f(q)}R^{f(q)}_{qb}\right].
\end{eqnarray}
Interestingly, the dependence on $R$ in Eqs. \eqref{eq:td_rho_glob} and \eqref{eq:td_rho_corr} cancel out such that $\dot{\rho}^{glob}$ is independent of $\hat{X}$. This is an important result because it means that the real-time pDMET equations of motion do not involve an inversion process, which can be numerically unstable and expensive.

This methodology presents a robust set of equations of motion that describe the coupled time evolution of the system at the mean-field and correlated levels.
In principle, it is only necessary to directly propagate the CAS coefficients, $C_M^A$, for all fragments and the mean-field density matrix, $\rho^{mf}$, using Eqs. \eqref{eq:time_dep_of_C} and  \eqref{eq:td_rho_mf}; all other time-dependent quantities can be calculated once these are known. However, to increase the numerical stability of the method, we additionally directly propagate the embedding orbitals, $R^A$, for all fragments, the natural orbitals, $U$, and the global density matrix $\rho^{glob}$ using Eqs. \eqref{eq:td_R}, \eqref{eq:td_U}, and \eqref{eq:td_rho_glob_final}, respectively. Full algorithmic details are provided in the SI. The working equations of motion have been implemented using in-house code that has been interfaced with the PySCF package\cite{sun2020recent}. Publicly available version of the code has been made available \cite{ rtdmetGit_link}.

\section{Results}

The performance of real time pDMET is examined by simulating electron dynamics in a set of 1-dimensional (1D) multi-impurity Anderson models (MIAMs) and single-impurity Anderson models (SIAMs). The first set of calculations involve suddenly changing the strength of the electron-electron interaction terms, while a second set of calculations additionally include a laser pulse. We conclude with a calculation on a 2-dimensional (2D) SIAM.

The SIAM models a system that consists of a single impurity with a local Coulomb interaction ($U$) and a nearest-neighbor hopping between all sites ($t$).

The MIAM extends this model to include multiple impurity sites, $N_{imp}$, that each has a local Coulomb interaction. The Hamiltonian that encompasses both the SIAM and MIAM is given by
\begin{equation}
    H = -t\sum_{<ij>\sigma}(a_{i\sigma}^\dag a_{j\sigma}+h.c.)+U\sum_{i\in imp}n_{i\uparrow}n_{i\downarrow}
\label{eq:Anderson_hamiltonian}
\end{equation}
The Coulomb interaction is described in terms of the occupation number operator $n_{i,\sigma}=a_{i\sigma}^\dag a_{i\sigma}$ on impurity site $i$. The sum $\sum\limits_{<ij>}$ implies a sum over only nearest-neighbors.
These models provide a useful and challenging benchmark platform for evaluating the feasibility and effectiveness of the real-time pDMET method\cite{titvinidze_strong-coupling_2015, eickhoff_strongly_2020}.

To generate non-equilibrium dynamics, we initialize the system in the non-interacting ground state of the SIAM/MIAM ($U=0$, $t=1$) by performing a static pDMET calculation using exact diagonalization to solve the embedding problem and subsequently propagate the dynamics using a Hamiltonian with a non-zero value of the interaction parameter ($U'\neq U$). Initializing the system at $U=0$ provides a useful benchmark for real-time pDMET because pDMET is exact in the non-interacting limit. To examine the accuracy of the dynamics, we benchmark the performance of real-time pDMET against TD-DMRG and TDHF theory. A discussion of the numerical implementation and full computational details are provided in the SI.

\begin{figure}[t!]
\centering
\includegraphics[scale=0.60]{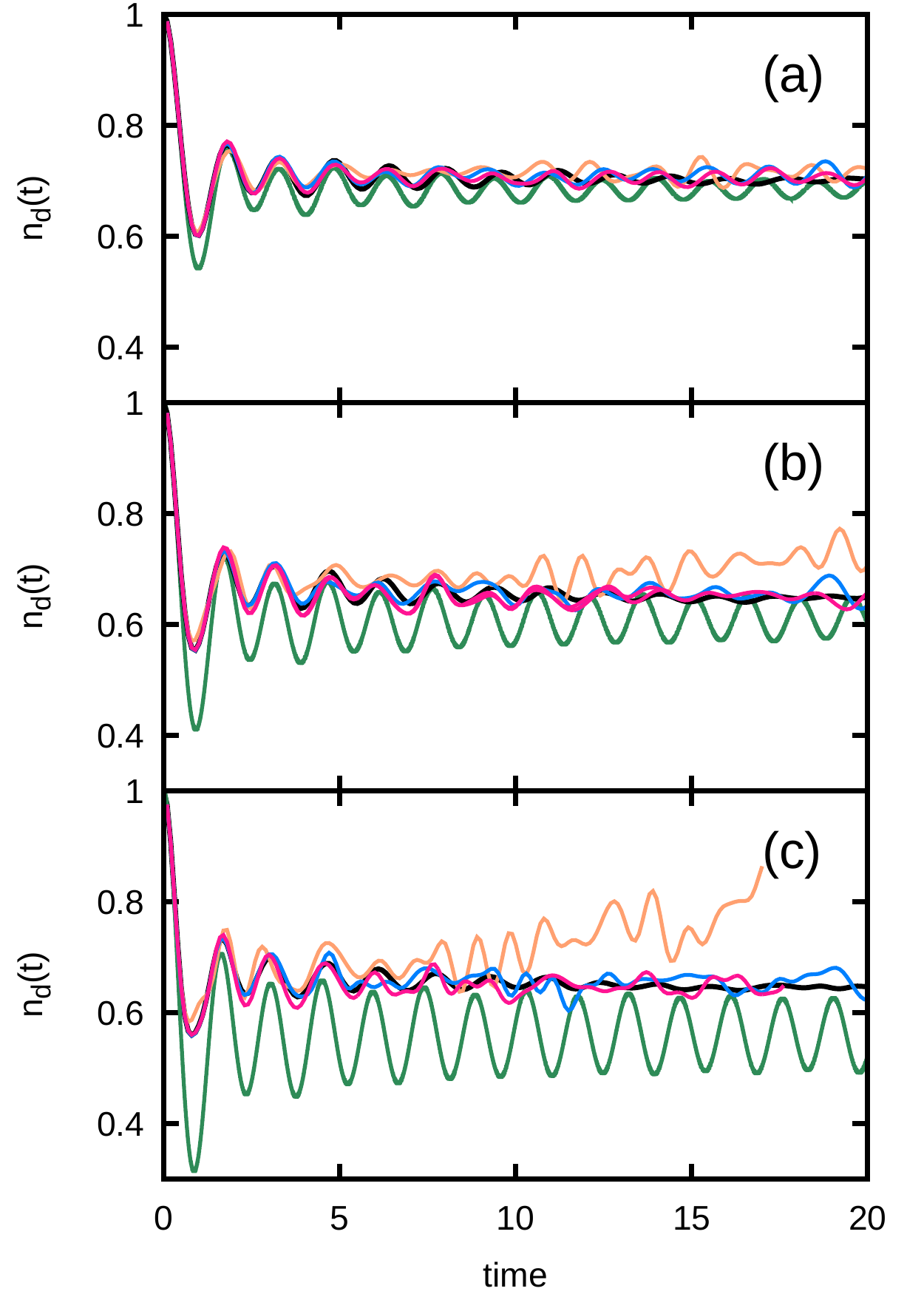}
\caption{The time dependent occupancy on the single impurity, $n_d(t)$, for the SIAM for which the strength of the Coulomb interaction is switched from $U=0$ to (a) $U'=2$, (b) $U'=3$ and (c) $U'=4$. Real-time pDMET results for $N_{sub}=2$  (orange), $N_{sub}=4$  (blue), $N_{sub}=5$ (pink) are compared to TD-DMRG (black) and TDHF (green). The total size of the system is $N=60$ with the single impurity located at $d=N/2$.} 
\label{fig:Nu1_60}
\end{figure}\

Figure \ref{fig:Nu1_60} presents results for a SIAM consisting of $N=60$ sites in which the initial Coulomb interaction is changed from $U=0$ to $U' = $ (a) 2, (b) 3, and (c) 4. Firstly, Figure \ref{fig:Nu1_60} shows that TDHF is able to reasonably capture the dynamics when correlation effects are weak ($U'=2$), but rapidly declines in accuracy as the importance of correlation effects increases with higher $U'$. In contrast, real-time pDMET always converges to the TD-DMRG results with increasing fragment size regardless of the magnitude of $U'$. While $N_{frag}=2$ is insufficient at higher $U'$, strongly correlated dynamics are fully reproduced with a fragment size as small as $N_{frag}=5$. This implies that a reduced active space of 5 subsystem and 5 bath orbitals for each fragment is sufficient to capture the correlated electron dynamics of this $N=60$ system.

\begin{figure}[t]
\centering
\includegraphics[scale=0.60]{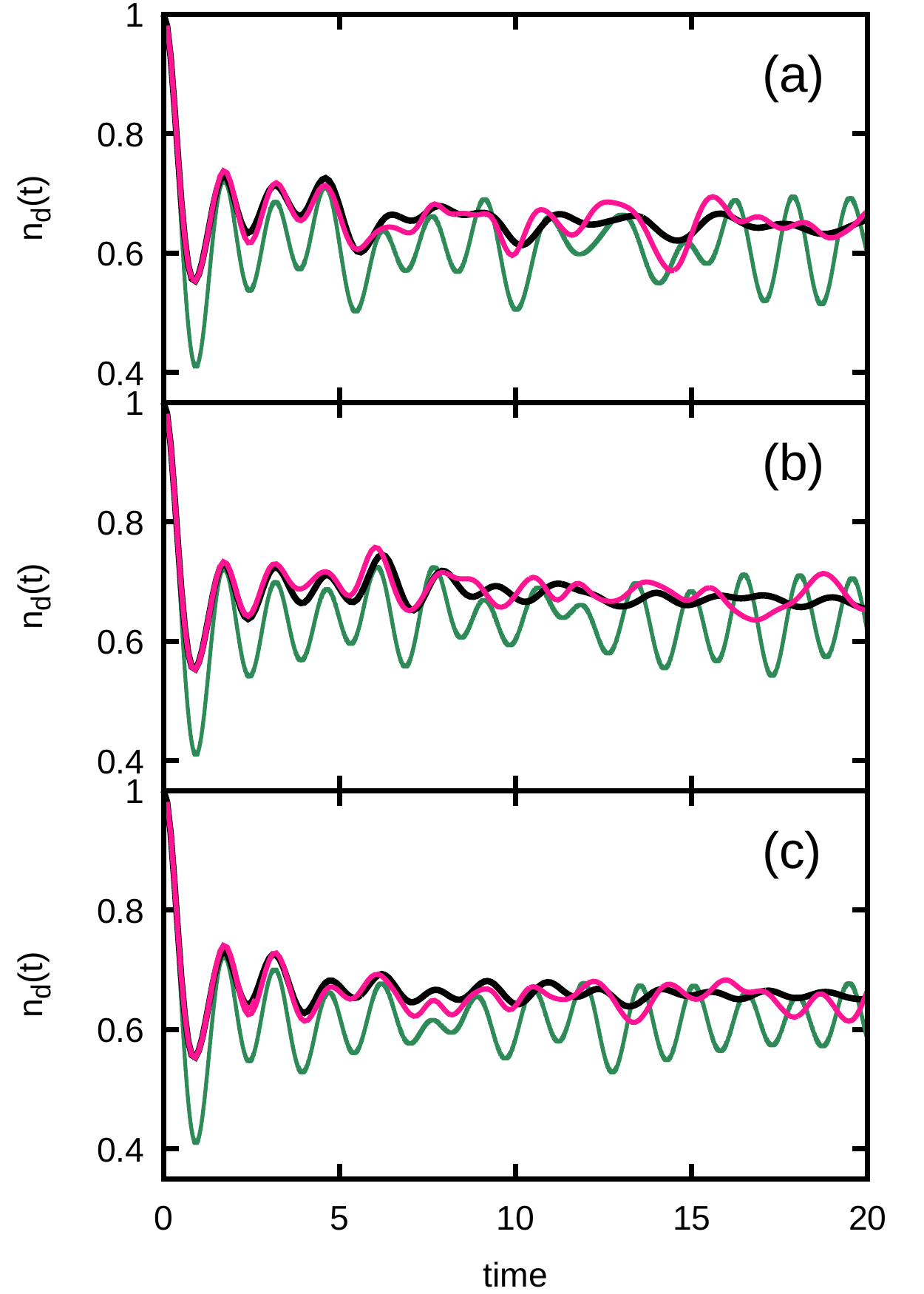}
\caption{The time dependent occupancy on one of the 6 impurities, $n_d(t)$, for the MIAM with (a) evenly distributed impurities and $d=65$, (b) unevenly distributed impurities and $d=67$, and (c) unevenly distributed impurities and $d=45$. Real-time pDMET results for $N_{sub}=5$ (pink) are compared to TD-DMRG (black) and TDHF (green). In all panels the strength of the Coulomb interaction is switched from $U=0$ to $U'=3$ and the size of the system is $N=120$.
}
\label{fig:Nu6_120}
\end{figure}

\begin{figure}[t!]
\centering
\includegraphics[scale=0.30]{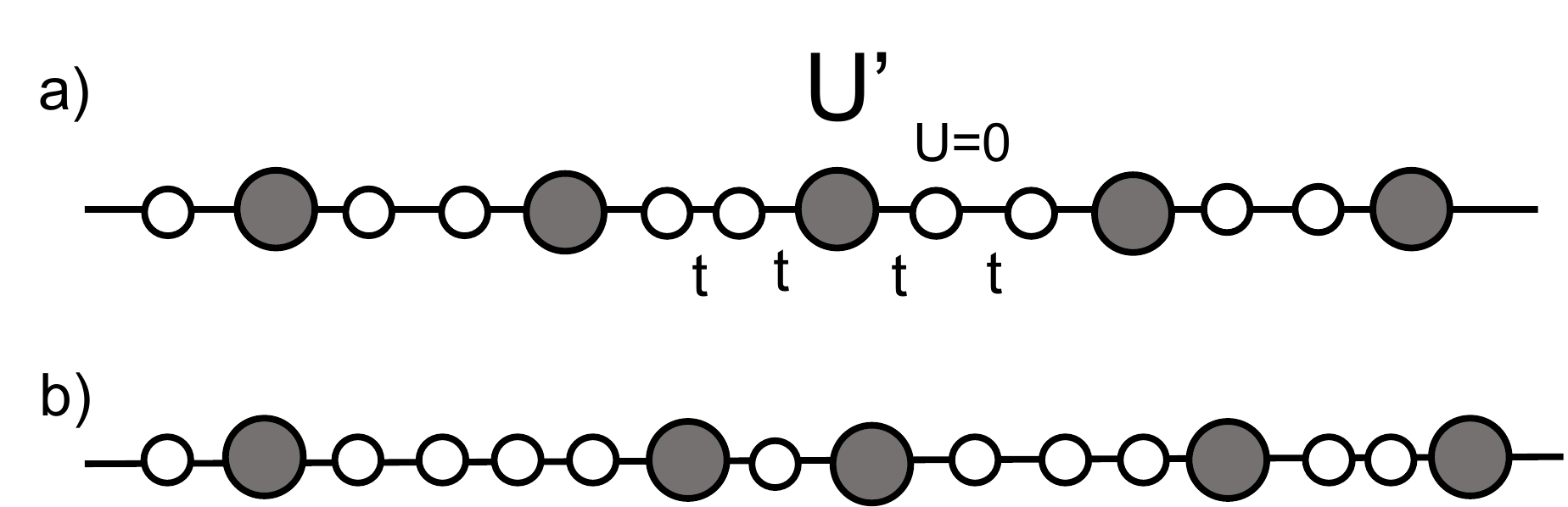}
\caption{Pictorial representation of the multi-impurity Anderson model with an (a) even and (b) uneven distribution of impurities (dark grey). Dynamics are initialized by suddenly changing the local Coulomb interaction from $U=0$ to a finite $U'$ only on the impurity sites; the interaction term remains $U=0$ on all other sites (white).} 
\label{fig:miam}
\end{figure}

Figure \ref{fig:Nu6_120} presents results for the more challenging case of the MIAM in which the  Coulomb interaction is changed from $U=0$ to $U'=3$. Figure \ref{fig:Nu6_120}(a) corresponds to a system in which the impurities are evenly distributed throughout the MIAM, while (b) and (c) present results for the time-dependent occupancy on different impurity sites in a MIAM with unevenly distributed impurities (Figure \ref{fig:miam}).

Analogous to the SIAM, TDHF fails to reproduce the dynamics in these strongly correlated systems, while real-time pDMET shows excellent agreement with the TD-DMRG results. 
The similar accuracy in parts (b) and (c) indicate that the multi-fragment formulation allows for an even treatment of the entire system.
The results for the unevenly distributed MIAM are especially encouraging as they indicate the ability for real-time pDMET to treat complex dynamics even in disordered systems.

\

\begin{figure}[t!]
\centering
\includegraphics[scale=0.55]{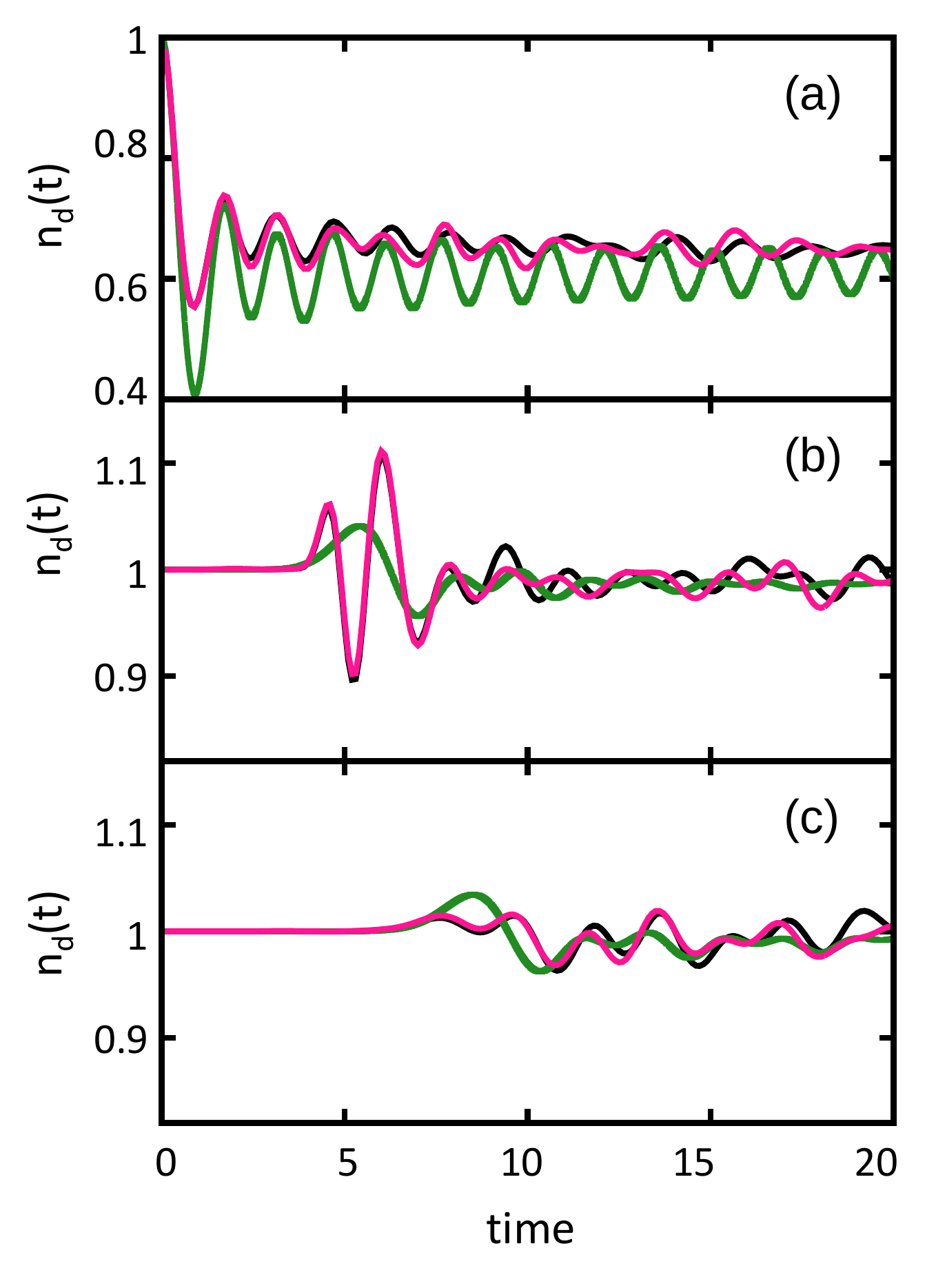}
\caption{ The time-dependent occupancy on various sites for a SIAM exposed to a laser pulse and a sudden change in the Coulomb interaction. (a) Site $d=30$, which corresponds to the impurity site containing the on-site Coulomb interaction. (b) Site $d=21$ located within the region exposed to the laser. (c) Site $d=15$ positioned away from both the laser region and the impurity site. Real-time pDMET results for $N_{sub}=5$ (pink) are compared to TD-DMRG (black) and TDHF (green).}
\label{fig:laser}
\end{figure}

We now turn our attention to results that additionally include an ultra-fast laser pulse to examine the behavior of rt-pDMET under a more complex non-equilibrium setting. Figure \ref{fig:laser} presents the time-dependent occupancy for the analogous $N=60$ SIAM system presented in Figure \ref{fig:Nu1_60}(b), but now including a laser pulse that is restricted to interact with only the central 20 sites of the system. The laser induces a time-dependence to the hopping terms of the Hamiltonian associated with these sites as described in detail in the SI. In addition to the laser pulse, we still include a sudden change from $U=0$ to $U'=3$ at $t=0$, such that the dynamics are initialized from an exact solution.

Figure \ref{fig:laser}(a) shows the occupancy on the impurity site and is comparable to Figure \ref{fig:Nu1_60}(b); the dynamics on this site are dominated by the sudden change in $U$ versus the interaction with the laser pulse. In comparison, Figure \ref{fig:laser}(b) shows the occupancy on a site removed from the central impurity but within the region interacting with the laser pulse. Here, real-time pDMET (pink) accurately captures the distinct change in occupancy of the laser-exposed site at the pumping time at $t=5$, which can not be described by TDHF (green) even on a qualitative level. Lastly, Figure \ref{fig:laser}(c) presents the occupancy on a site outside of the interaction region with the laser pulse. Real-time pDMET shows excellent agreement with TD-DMRG in capturing the delayed dynamics that don't appear until $t\sim7.5$ associated with electron flow from the central impurity site and the region interacting with the laser.

\

\begin{figure}[t!]
\centering
\includegraphics[scale=0.45]{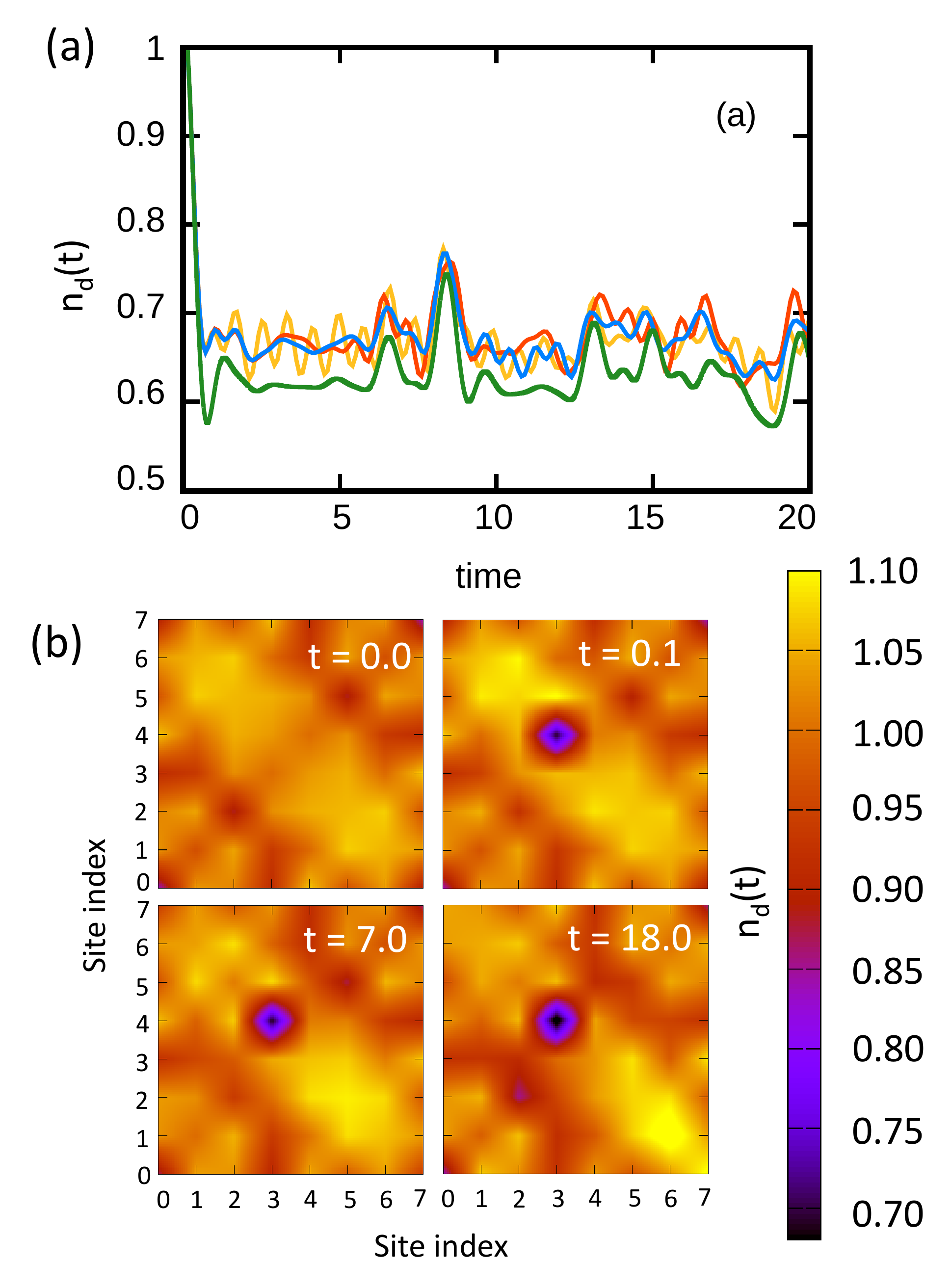}
\caption{ The time dependent occupancy on sites within 2D SIAM of total size $N=64$, where electron dynamics are driven by Coulomb interaction change from $U=0$ to $U'=3$. (a) Site $d=(3,4)$, which corresponds to the impurity site containing the on-site Coulomb interaction. TDHF (green) results are compared to the real-time pDMET results for three fragment sizes: $N_{sub}=1\times1$ (yellow), $N_{sub}=1\times2$ (orange), $N_{sub}=1\times4$ (blue). (b) Two-dimensional electron density distribution across the full system at four different points of time during the simulation.} 
\label{fig:U3_2D}
\end{figure}\

We conclude by illustrating the flexibility of real-time pDMET towards more realistic systems beyond what can be easily treated by TD-DMRG. Specifically, we examine the dynamics in a 2D SIAM initialized by a sudden switch in the value of $U$ from 0 to 3 on one of the central sites. TD-DMRG is poorly suited to treat multi-dimensional systems due to the structure of the matrix-product state \emph{Ans\"{a}tze} for the wavefunction.\cite{Bai20} Figure \ref{fig:U3_2D}(a) shows the time-dependent occupation on the impurity that contains the on-site Coulomb interaction. Real-time pDMET demonstrates a rapid convergence of the dynamics with an increase in fragment size. The oscillatory behaviour observed in the $N_{sub}=1\times1$ results  (yellow) is eliminated by a minimum increase in fragment size to $N_{sub}=1\times2$ (orange). Sequential increase of the fragment size to $N_{sub}=1\times4$ (blue) demonstrates convergence of the dynamics that is distinct from the TDHF results (green). 
 
 Figure \ref{fig:U3_2D}(b) shows the electron density in the full system at four representative times throughout the simulation. The rapid change in the occupation of the site with the Coulomb interaction is observed immediately after the change in the Hamiltonian. Sequential redistribution of electron density is observed from the upper left corner to the bottom right corner of the system as a function of time associated with a response to the sudden change in $U$ and reflections off of the boundaries of the system. This electron flow is what leads to the recurrence peak around $t=8$ in Figure \ref{fig:U3_2D}(a).

\section{Conclusion}

In this work, we present the development of a real-time extension of projected density matrix embedding theory (real-time pDMET). The analytic matching condition used to obtain self-consistency in static pDMET compared to the correlation potential utilized in conventional DMET removes previous theoretical limitations associated with real-time extensions of DMET. This allows for a multi-fragmented formulation of the dynamics, analogous to static pDMET and DMET.

The real-time pDMET method was benchmarked for non-equilibrium electron dynamics in both single-impurity and multiple-impurity  1- and 2-dimensional Anderson models (SIAM and MIAM). Dynamics were initialized through a sudden change in the electron interaction term and through an external laser pulse. In all cases, real-time pDMET converged rapidly to numerically exact TD-DMRG results with increasing fragment size. Most notably real-time pDMET was able to accurately capture the dynamics across all regions of space in a disordered MIAM and exhibited rapid convergence of the dynamics in a 2D SIAM, for which TD-DMRG is poorly suited.

Overall, the multi-fragmented formulation of real-time pDMET provides an efficient and flexible framework to investigate time-dependent properties of strongly correlated systems. The embedding framework is not limited by the dimensionality of the system and allows for the future incorporation of other wavefunction \emph{Ans\"{a}tze}, such as coupled-cluster or matrix-product state wavefunctions, as has been done in the context of static calculations. Future development will involve an unrestricted formulation of the method, which will allow for higher accuracy and explicit treatment of spin dynamics.

\section{Supplementary Material}

The supplementary material contains a full derivation of the working dynamical equations for the real-time pDMET method along with a discussion of the numerical implementation and full computational details. Publicly available version of the real-time pDMET code used to generate the results in this manuscripts has been made available\cite{ rtdmetGit_link}.

\begin{acknowledgments}
This work was partially supported by a Georgia Tech Seed grant. Computational resources were supported in part through research cyberinfrastructure resources and services provided by the Partnership for an Advanced Computing Environment (PACE) at the Georgia Institute of Technology, Atlanta, Georgia, USA.
\end{acknowledgments}


\begin{titlepage}
  \centering
  \vskip 60pt
  \LARGE Supporting Information \par
  \vskip 3em
  \large A multi-fragment real-time extension of projected density matrix embedding theory: Non-equilibrium electron dynamics in extended  systems \par
  \vskip 1.5em
  \large D. Yehorova, Joshua S. Kretchmer \par
  \vskip 1.5em
  
\end{titlepage}

\newpage
\input{SI.tex}
\end{document}

%% file: SI.tex












\section*{Detailed derivations of the equations of motion}

Here we provide complete derivations for the real-time pDMET equations of motion defined in the main text. The notation used throughout the following derivations is consistent with the notation used in the methods section of main text. The site basis is indexed by $p, q, r, s$; the embedding basis is indexed by $a, b, c$ and $d$; and the natural orbital basis is indexed by $\mu$ and $\nu$.

Two-electron integrals are defined using chemist notation,
\begin{equation}
V_{pqrs}=\int\int\phi_p^*(r_1)\phi_r^*(r_2)\frac{1}{r_{12}}\phi_q(r_1)\phi_s(r_2)dr_1dr_2.
\end{equation}
The 1- and 2-particle reduced density matrices are defined as:
\begin{eqnarray}
   {\rho}_{pq}&=&\langle\Psi|\hat{a}^\dag_{q\sigma}\hat{a}_{p\sigma}|\Psi\rangle
   \label{eq:SI_1RDM_def}\\
\Gamma_{pqrs}&=&\sum_{\sigma\tau}\langle\Psi|a_{p\sigma}^{\dag}a_{r\tau}^{\dag}a_{s\tau}a_{q\sigma}|\Psi\rangle,
\label{eq:SI_2RDM_def}
\end{eqnarray}
respectively.
Both the embedding and natural orbitals define a rotation between the site and the corresponding basis. 
\begin{equation}
\begin{aligned}
|a\rangle=\sum_{p}R_{pa}|p\rangle, && R_{pa}=\langle p|a\rangle
\label{eq:SI_R_def}
\end{aligned}
\end{equation}
\begin{equation}
\begin{aligned}
|\mu\rangle=\sum_{p}U_{p\mu}|p\rangle, && U_{p\mu}=\langle p|\mu\rangle
\label{eq:SI_U_def}
\end{aligned}
\end{equation}

\subsection{Derivation of $X$ and $G$ matrices}

The operators $\hat{X}$ and $\hat{G}$ describe time evolution of the embedding and natural orbitals respectively. The following discussion focuses on the derivation of the matrix elements of $X$, Eq. (20) in the main text; the derivation of the matrix elements of $G$, Eq. (23) in the main text, follows completely analogous steps.

The derivation of $X$ is based on two previously described conditions: the pDMET condition that defines the relationship between $\rho^{mf}$ and $R$ and the initial definition of $X$ as a propagator of the embedding orbitals.
 \begin{eqnarray}
    \sum_{q \in env}^{N-N_{A}}\rho^{mf}_{pq}R^{A}_{qa} &=& \lambda^{A}_{a}R^{A}_{pa}
    \label{eq:SI_RDM_mf_and_R}\\
    i\dot{R}_{pa}^{A}&=&\sum_{q}X_{pq}^{A}R_{qa}^{A}\label{eq:SI_td_R}
\end{eqnarray}
Each fragment, $A$, has a corresponding set of embedding orbitals, which implies that there is a unique $\hat{X}^A$ for each fragment. However, the explicit reference to fragment $A$ will be suppressed in the following derivation for convenience.

The derivation of the non-subsystem elements of $X$ begins by taking the time-derivative of Eq. \eqref{eq:SI_RDM_mf_and_R}:

\begin{equation}
    \sum_{q \in env}\left(\dot{\rho}^{mf}_{pq}R_{qa}+\rho^{mf}_{pq}\dot{R}_{qa}\right)=\dot{\lambda}_aR_{pa}+\lambda_a\dot{R}_{pa}
\label{eq:SI_td_mf_condition}
\end{equation}
The right side of the equation can be rewritten by explicitly defining $\dot{\lambda}_a$ using Eq. \eqref{eq:SI_RDM_mf_and_R}:
\begin{eqnarray}
    \lambda_a&=&\sum_{qr \in env}R_{qa}^*\rho^{mf}_{qr}R_{ra}\\
    \dot{\lambda}_a&=&\sum_{qr \in env}\left[\dot{R}_{qa}^*\rho^{mf}_{qr}R_{ra}+R_{qa}^*\rho^{mf}_{qr}\dot{R}_{ra}+R_{qa}^*\dot{\rho}^{mf}_{qr}R_{ra}\right]
\label{eq:SI_td_lambda_with_RDM}
\end{eqnarray}
Using the definition of the matrix $X$ stated in Eq. \eqref{eq:SI_td_R}, Eq. \eqref{eq:SI_td_lambda_with_RDM}, and the pDMET condition Eq. \eqref{eq:SI_RDM_mf_and_R} yields

\begin{widetext}
\begin{equation}
   i\dot{\lambda}_a=\sum_{q \in env}\sum_s\left(-R^*_{sa}X_{sq}\lambda_aR_{qa}\right)+\sum_{r \in env}\sum_{s}\left(\lambda_aR_{ra}^{*}X_{rs}R_{sa}\right) +\sum_{qr \in env}\left(R_{qa}^{*}i\dot{\rho}_{qr}^{mf}R_{ra}\right)\\
\label{eq:SI_td_lambda_with_X_and_RDM}
\end{equation}
\end{widetext}
This expression can be further simplified by realizing that the subsystem orbitals are time-independent; this condition implies that $X_{sq}=0$ if either $s$ or $r$ is a subsystem orbital of the fragment. Therefore, both sums can be restricted to only the environment orbitals. This simplification allows for the rearrangement of indices such that the first two terms cancel, reducing the definition of $i\dot{\lambda}_a$ in Eq. \eqref{eq:SI_td_lambda_with_X_and_RDM} to only the last term.

The matrix $X$ is similarly introduced into the left side of Eq. \eqref{eq:SI_td_mf_condition} by using Eqs. \eqref{eq:SI_RDM_mf_and_R} and \eqref{eq:SI_td_lambda_with_RDM}, such that Eq. \eqref{eq:SI_td_mf_condition} becomes
\begin{widetext}
\begin{equation}
    \sum_{q \in env}\dot{\rho}^{mf}_{pq}R_{qa}+\sum_{q \in env}\sum_r\rho^{mf}X_{qr}R_{ra}=\sum_{qr \in env}\left(R_{qa}^{*}i\dot{\rho}_{qr}^{mf}R_{ra}\right)R_{pa}+\lambda_a\sum_rX_{pr}R_{ra}
    \label{eq:SI_x_pre_multiplication_by_b}
\end{equation}
\end{widetext}

Multiplying on the left by $\langle b|$ using the definition \eqref{eq:SI_R_def}, and then further simplifying using the complex conjugate of Eq. \ref{eq:SI_RDM_mf_and_R}, the fact that $\rho^{mf}$ is Hermitian, $\lambda_a$ is real, and the unitarity of the matrix $R$ yields
\begin{equation}
\begin{aligned}
    \sum_{pq \in env}R_{pb}^*i\dot{\rho}^{mf}_{pq}R_{qa}+\sum_{q\in env}\sum_r\lambda_bR_{pb}^*X_{qr}R_{ra}=\\
    \sum_{qr \in env}\left(R_{qa}^{*}i\dot{\rho}_{qr}^{mf}R_{ra}\right)\delta_{ba}+\sum_{p\in env}\sum_r\lambda_aR_{pb}^*X_{pr}R_{ra}
    \label{eq:SI_x_post_multiplication_by_b}
\end{aligned}
\end{equation}
All summations can now be restricted to the environment space by an analogous argument used in Eq. \eqref{eq:SI_td_lambda_with_X_and_RDM}, such that
\begin{eqnarray}
    \sum_{pq \in env}R_{pb}^*i\dot{\rho}^{mf}_{pq}R_{qa}+\sum_{qr\in env}\lambda_bR_{pb}^*X_{qr}R_{ra}=\nonumber\\
    \sum_{qr \in env}\left(R_{qa}^{*}i\dot{\rho}_{qr}^{mf}R_{ra}\right)\delta_{ba}+\sum_{pr\in env}\lambda_aR_{pb}^*X_{pr}R_{ra}
\end{eqnarray}
This equation can be simplified using that the matrix $X$ in the embedding space is defined as
\begin{eqnarray}
X_{ba}&=&\sum_{pr\in env}R_{pb}^*X_{pr}R_{ra}
\end{eqnarray}
and that due to the redundancy of intraspace rotations, the nonzero terms of $X$ correspond to $a\neq b$. Therefore, the final expression for the nonzero elements correspond to

\begin{equation}
   X_{ba}=\sum_{pq \in env}\frac{R_{pb}^*i\dot{\rho}^{mf}_{pq}R_{qa}}{\lambda_a-\lambda_b},
\label{eq:SI_td_lambda_with_X_and_RDM_final}
\end{equation}
which corresponds to Eq. (20) in the main text.

The derivation of the operator governing the time-dependence for the natural orbital is completely analogous where equations \eqref{eq:SI_RDM_mf_and_R} and \eqref{eq:SI_td_R} are replaced by the corresponding relationships in natural orbital space and all operations are carried out in the global domain. 
 \begin{eqnarray}
    \sum_{q}\rho^{glob}_{pq}U_{q\mu} &=& \lambda_{\mu}U_{p\mu}
    \label{eq:SI_RDM_glob_and_U}\\
    i\dot{U}_{p\mu}&=&\sum_{q}G_{pq}U_{q\mu}\label{eq:SI_td_U}
\end{eqnarray}
The final expression of the matrix elements of the operator $G$ is: 
\begin{equation} \label{eq:SI_G_def}
    G_{\mu\nu}=\sum_{pq}\frac{U_{p\mu}^{*}i\dot{\rho}^{glob}_{pq}U_{q\nu}}{\lambda_{\nu}-\lambda_{\mu}},
\end{equation}
which corresponds to Eq. (23) in the main text.

\subsection{Derivation of $\dot{\rho}^{mf}$}
The time-dependence of $\rho^{mf}$ is directly obtained from the dynamics of the natural orbitals $U$. 
This is observed after taking the derivative of the pDMET condition \eqref{eq:SI_rho_mf_and_rho_glob} which ensures the consistency between the mean-field representation and global domain of the system. 
\begin{eqnarray}
    \rho^{mf}_{pq}&=& 2 \sum_{\mu}^{N_{occ}}U_{p\mu}U_{q\mu}^{*} 
    \label{eq:SI_rho_mf_and_rho_glob}\\
    i\dot{\rho}^{mf}_{pq} &=& 2 \sum_{\mu}^{N_{occ}}\left(i\dot{U}_{p\mu}U_{q\mu}^{*}+U_{p\mu}i\dot{U}_{q\mu}^{*}\right)
    \label{eq:SI_td_rho_mf}
\end{eqnarray}
The time dependence of $U$ is defined in terms of the operator $\hat{G}$, Eq. \eqref{eq:SI_td_U}. Therefore, Eq. \eqref{eq:SI_td_rho_mf} can be rewritten as a commutator of $G$ and $\rho^{mf}$ in the site basis,
\begin{eqnarray}
    i\dot{\rho}^{mf}_{pq} &=& 2 \sum_{\mu}^{N_{occ}}\sum_{r}\left(G_{pr}{U}_{r\mu}U_{q\mu}^{*}-U_{p\mu}{U}_{r\mu}^{*}G_{rp}\right)\nonumber\\
                    &=&\sum_{r}\left(G_{pr}\rho^{mf}_{rq}-\rho^{mf}_{pr}G_{rp}\right),
    \label{eq:SI_td_rho_mf_final}
\end{eqnarray}
which corresponds to Eq. (22) in the main text.

\subsection{Derivation of $\dot{\rho}^{emb}$}

The time-dependence of the correlated 1-RDM associated with each fragment, $(\rho_{ab}^{emb})^A$, is necessary for the derivation of $\dot{\rho}^{glob}$ seen below. This 1-RDM is obtained from the fragment wavefunction as
\begin{equation}
(\rho_{ab}^{emb})^A=\langle \Psi^A|\sum_{\sigma}a^{\dag A}_{b\sigma}a^A_{a\sigma}|\Psi^A\rangle
\end{equation}
The time derivative of this expression results in 4 terms, where the explicit dependence on the fragment $A$ is suppressed for convenience
\begin{equation}
\begin{aligned}
i\dot{\rho}_{ab}=i\langle\dot{ \Psi}|\sum_{\sigma}a^{\dag}_{b\sigma}a_{a\sigma}|\Psi\rangle+i\langle \Psi|\sum_{\sigma}a^{\dag}_{b\sigma}a_{a\sigma}|\dot{\Psi}\rangle\\
+\langle{ \Psi}|\sum_{\sigma}i\dot{a}^{\dag}_{b\sigma}a_{a\sigma}|\Psi\rangle+\langle{ \Psi}|\sum_{\sigma}ia^{\dag}_{b\sigma}\dot{a}_{a\sigma}|\Psi\rangle
\end{aligned}
\end{equation}
The first two terms are redefined as a commutator using the time-dependent Schr\"{o}dinger equation (TDSE); the second two terms can be written in terms of the time-dependence of the embedding orbitals by introducing the transformation of the creation and annihilation operators from the site to the embedding basis, 

\begin{equation}
\begin{aligned}
a_{a\sigma}^{\dag}=\sum_pR_{pa}a_{p\sigma}^{\dag}, && a_{a\sigma}=\sum_pR^*_{pa}a_{p\sigma}
\end{aligned}
\end{equation}

\begin{equation}
\begin{aligned}
i\dot{\rho_{ab}}=\langle\Psi|\left[\sum_{\sigma}a^{\dag}_{b\sigma}a_{a\sigma}, H\right]|\Psi\rangle+\\
\langle\Psi|\sum_{\sigma p}\left(i\dot{R}_{pb}a^{\dag}_{p\sigma}a_{a\sigma} + a^{\dag}_{b\sigma}i\dot{R}_{pa}^*a_{p\sigma}\right)|\Psi\rangle\\
\end{aligned}
\end{equation}
This is further simplified by introducing the matrix $X$ using Eq. \eqref{eq:SI_td_R} and solving for the commutator using the definitions of the 1- and 2-RDMs, Eqs. \eqref{eq:SI_1RDM_def} and \eqref{eq:SI_2RDM_def}. 
\begin{widetext}
\begin{eqnarray}
   i\dot{\rho_{ab}} &=&\langle\Psi|\left[\sum_{\sigma}a^{\dag}_{b\sigma}a_{a\sigma}, H\right]|\Psi\rangle+\sum_c\left(\rho_{ac}X_{cb}-X_{ac}\rho_{cb}\right)\\
&=&\sum_c\left(h_{ac}\rho_{cb}-\rho_{ac}h_{cb}\right)+\sum_{cde}\left(V_{acde}\Gamma_{bcde}-V_{cbde}\Gamma_{cade}\right)+\sum_c\left(\rho_{ac}X_{cb}-X_{ac}\rho_{cb}\right)\label{eq:SI_dt_corr1DM}
\end{eqnarray}

\end{widetext}
The first line of Eq. \eqref{eq:SI_dt_corr1DM} can be written in terms of a generalized Fock matrix in a completely analogous form to what is done both in TD-CASSCF\cite{sato_time-dependent_2013} and static CASSCF theory\cite{helgaker_recent_2012}, with the one caveat that the orbitals are complex-valued.
\begin{eqnarray}
i\dot{\rho}_{ab}&=&F_{ba}-F_{ab}^*+\sum_c\left(\rho_{ac}X_{cb}-X_{ac}\rho_{cb}\right)\\&\equiv&i\dot{\tilde{\rho}}_{ab}+\sum_c\left(\rho_{ac}X_{cb}-X_{ac}\rho_{cb}\right)
\label{eq:SI_dt_corrRDM_withF}
\end{eqnarray}
This corresponds to Eq. (25) in the main text. 

The generalized Fock matrix is given by
\begin{eqnarray}
F_{ba}&=&\sum_ch_{ac}\rho_{cb}+\sum_{cde}V_{acde}\Gamma_{bcde}.
\end{eqnarray}
The expression for the generalized Fock matrix can be further simplified using the specific form of the 1- and 2-RDMS for the CAS-like wavefunction for each fragment. Specifically, $\rho_{cb}=0$ and $\Gamma_{bcde}=0$ if any index corresponds to a virtual embedding orbital. Furthermore, $\rho_{cb}=2\delta_{cb}$ if either $c$ or $b$ corresponds to a core embedding orbital, $\Gamma_{bcde}=4\delta_{bc}\delta_{de}-\delta_{be}\delta_{cd}$ if all indices correspond to a core orbital, $\Gamma_{bcde}=2\rho_{ed}\delta_{bc}$ if only $b$ and $c$ correspond to a core orbital, and $\Gamma_{bcde}=-\rho_{cd}\delta_{be}$ if only $b$ and $e$ correspond to a core orbital. Using these simplifications and the symmetries of the 1- and 2-RDMs, the generalized Fock matrix can be written as
\begin{widetext}
\begin{equation}
F_{ba}=
\begin{cases}
2^IF_{ab}+2^AF_{ab} & \mbox{if $b$ is a core orbital}\\
\sum\limits_{c\in sub/bath}\,^IF_{ac}\rho_{cb}+\sum\limits_{cde\in sub/bath}V_{acde}\Gamma_{bcde} & \mbox{if $b$ is a subsystem or bath orbital}\\
0 & \mbox{if $b$ is a virtual orbital},\\
\end{cases}
\label{eq:SI_mod_F}
\end{equation}
\end{widetext}
where the summations in the second line of Eq. \eqref{eq:SI_mod_F} are over the subsystem and bath orbitals. Finally, the inactive and active Fock matrices in Eq. \eqref{eq:SI_mod_F} are given by\cite{helgaker_recent_2012}
\begin{eqnarray}
^IF_{ab}&=&h_{ab}+\sum_{c\in core}\left(2V_{abcc}-V_{accb}\right)\\
^AF_{ab}&=&\sum_{cd\in sub/bath}\rho_{cd}\left(V_{abdc}-\frac{1}{2}V_{acdb}\right)\label{eq:SI_act_F}
\end{eqnarray}

\subsection{Derivation of $\dot{\rho}^{glob}$}

The explicit time dependence of $\rho^{glob}$ is necessary to define the operator $\hat{G}$ that governs the mean-field dynamics, Eq. \eqref{eq:SI_G_def}. The global 1-RDM is defined through the democratic partitioning of the correlated 1-RDMs from the separate fragment calculations, such that

\begin{equation}
\begin{aligned}
  {\rho}^{glob}_{pq}= \frac{1}{2}\sum_{ab}\left[R_{pa}^{f(p)}{(\rho}_{ab}^{emb})^{f(p)}R^{*f(p)}_{qb}\right.\\
  \left.+ R_{pa}^{f(q)}({\rho}^{emb}_{ab})^{f(q)}R^{*f(q)}_{qb}\right]
\end{aligned}
\end{equation}
To simplify the notation in the following steps, $\rho^{emb}$ will simply be referred to as $\rho$ and the objects that belong to a fragment $f(q)$ will be referred with a singular superscript $q$. Using this notation, the time dependence of $\rho^{glob}$ involves the following six terms,
\begin{equation}
\begin{aligned}
    i\dot{\rho}^{glob}_{pq}= \frac{1}{2}\sum_{ab}&\left[i\dot{R}_{pa}^{p}{\rho}_{ab}^{p}R^{*p}_{qb} +i{R}_{pa}^{p}{\rho}_{ab}^{p}\dot{R}^{*p}_{qb}\right.\\
     &+\left.i\dot{R}_{pa}^{q}{\rho}_{ab}^{q}R^{*q}_{qb}
 +i{R}_{pa}^{q}{\rho}_{ab}^{q}\dot{R}^{*q}_{qb}\right.\\
     & + \left.R_{pa}^{p}i\dot{\rho}_{ab}^{p}R^{*p}_{qb}+ R_{pa}^{q}i\dot{\rho}_{ab}^{q}R^{*q}_{qb}\right].
\end{aligned}
\end{equation}
Inserting the definition of $\dot{R}$, Eq. \eqref{eq:SI_td_R}, and $\dot{\rho}^{emb}$, Eq. \eqref{eq:SI_dt_corrRDM_withF}, yields 

\begin{widetext}
\begin{equation}
\begin{aligned}
    i\dot{\rho}^{glob}_{pq}= \frac{1}{2}\sum_{abc}&\left[{R}_{pc}^{p}X_{ca}^p{\rho}_{ab}^pR^{*p}_{qb} -{R}_{pa}^{p}{\rho}_{ab}^{p}X_{bc}^p{R}^{*p}_{qc}+{R}_{pc}^{q}X_{ca}^q{\rho}_{ab}^{q}R^{*q}_{qb}
 -{R}_{pa}^{q}{\rho}_{ab}^{q}X_{bc}^q{R}^{*q}_{qc}\right]\\
     +\frac{1}{2}\sum_{ab}&\left[R_{pa}^{p}i\dot{\tilde{\rho}}_{ab}^{p}R^{*p}_{qb}+ R_{pa}^{q}i\dot{\tilde{\rho}}_{ab}^{q}R^{*q}_{qb}\right]+\frac{1}{2}\sum_{abc}\left[{R}_{pa}^{p}{\rho}_{ac}^{p}X_{cb}^p{R}^{*p}_{qb}-{R}_{pa}^{p}X_{ac}^p{\rho}_{cb}^pR^{*p}_{qb}\right.\\
     &+\left.{R}_{pa}^{q}{\rho}_{ac}^{q}X_{cb}^q{R}^{*q}_{qb}-{R}_{pa}^{q}X_{ac}^q{\rho}_{cb}^qR^{*q}_{qb}
 \right]\\
\end{aligned}
\end{equation}
\end{widetext}
All terms that contain a dependence on the matrix $X$ cancel out above, such that the time-dependence of the global density matrix only depends on the generalized Fock matrices,

\begin{equation}
\begin{aligned}
i\dot{\rho}^{glob}_{pq}=& \frac{1}{2}\sum_{ab}\left[R_{pa}^{p}i\dot{\tilde{\rho}}_{ab}^{p}R^{*p}_{qb}+ R_{pa}^{q}i\dot{\tilde{\rho}}_{ab}^{q}R^{*q}_{qb}\right]\\
=& \frac{1}{2}\sum_{ab}\left[R_{pa}^{p}i(F_{ba}-F_{ab}^*)^{p}R^{*p}_{qb}\right.\\
+& \left.R_{pa}^{q}i(F_{ba}-F_{ab}^*)^{q}R^{*q}_{qb}\right].
\label{eq:SI_td_rho_glob}
\end{aligned}
\end{equation}
This corresponds to Eq. (26) in the main text.

\section*{Computational Details}

 \subsection{System parameters}

In this work we simulate a 1-dimensional (1D) single-impurity Anderson model (SIAM) with and without a laser pulse, a set of 1D multi-impurity Anderson models (MIAM), and a 2-dimensional SIAM. The Hamiltonian that encompasses both the SIAM and MIAM in 1D and 2D, but without the laser pulse is given by
\begin{equation}
    H = -t\sum_{<ij>\sigma}(a_{i\sigma}^\dag a_{j\sigma}+h.c.)+U\sum_{i\in imp}n_{i\uparrow}n_{i\downarrow}
\end{equation}
which corresponds to Eq. (27) in the main text. 

In every calculation throughout the manuscript, the initial state of the system corresponds to the ground-state of the non-interacting system, where $U=0$ and $t=1$. As described in the main text, non-equilibrium dynamics are then initialized by running with a new value of the interaction term $U'\neq U$. The hopping term is fixed at $t=1$ between all sites in all systems, even during the dynamics.

The 1D SIAM consists of $N=60$ sites where the impurity is located at $i=\frac{N}{2}$. This system was used in Figure 2 in the main text.

We examine two types of MIAMs, both consisting of $N=120$ sites with $N_{imp}=6$ impurities. The first type has the impurities evenly distributed amongst the sites, located at sites $i=\{41, 49, 57, 65, 73, 81\}$. This corresponds to Figure 3(a) in the main text. The second type has the impurities unevenly distributed amongst the sites, located at sites $i=\{45, 51, 56, 67, 74, 80\}$. This corresponds to Figures 3(b) and (c) in the main text.

We also simulate the same 1D SIAM consisting of $N=60$ sites, but with the addition of a laser pulse that is restricted to interact with only the central 20 sites of the system. The influence of the external electric field is included into the Hamiltonian via the Peierls substitution in the hopping terms between sites 20 to 40.
\begin{equation}\label{eq:Peierls_sub}
a^\dag_{i,\sigma}a_{i+1,\sigma}+h.c. \xrightarrow{} e^{iA(t)}a^\dag_{i,\sigma}a_{i+1,\sigma}+h.c.,
\end{equation}
where $A(t)$ is the vector potential. In this work we choose $A(t)$ to have the common form of 
\begin{equation}\label{eq:laser}
A(t)=A_0e^{-(t-t_0)^2/2t^2_d}cos[\omega_0(t-t_0)],
\end{equation}
which involves an oscillating field of amplitude $A_0$ and central frequency $\omega_0$ with a Gaussian envelope centered in time at $t_0$ of width $t_d$. The parameters for the vector potential were chosen in agreement with previous work\cite{shao_2020}, such that $A_o=0.3$, $\omega=4.0$, $t_o=5$, and $t_d=0.5$. In addition to the laser pulse, we still include a sudden change from $U=0$ to $U'\neq U$ at $t=0$, such that the dynamics are initialized from an exact solution. This corresponds to Figure 5 in the main text.

Lastly, we also simulate a 2D SIAM without a laser pulse consisting of 8 sites in both the $x$ and $y$ directions (total N=64 sites). The impurity is located at ($x=3$, $y=4$). This corresponds to Figure 6 in the main text.

\subsection{Numerical implementation of real-time pDMET}

 The initial state for every real-time pDMET calculation in the manuscript is obtained from the ground-state solution of a static pDMET calculation of the corresponding non-interacting system; the interaction term $U=0$ for all sites. The static pDMET calculation is initialized using the 1-RDM obtained from a restricted Hartree-Fock calculation. Exact diagonalization is used as the solver for each fragment embedding problem. The static pDMET calculation converges in a single-step as the restricted Hartree-Fock calculation is exact for the non-interacting system and therefore provides the exact set of bath states. The static pDMET calculation outputs a mean-field 1-RDM, $\rho^{mf}$, (Eq. (3) in the main text), a set of embedding orbitals for each fragment, $R^{emb}$, (Eq. (4) in the main text), a set of CI coefficients for each fragment, $C$, (Eq. (5) in the main text), a global 1-RDM, $\rho^{global}$, (Eq. (7) in the main text), and a set of natural orbitals, $U$, (Eq. (9) in the main text). These are used as the inputs for the subsequent real-time pDMET calculation.

The real-time pDMET calculation involves the simultaneous propagation of the CAS-like wavefunction for each fragment and the mean-field 1-RDM. Practically this implies that we directly propagate the expansion coefficients, the embedding orbitals, and the mean-field 1-RDM as a function of time; the equations of motion for these three quantities are given by Eqs. (19), (18), and (22) in the main text, respectively. In addition, we also directly propagate the global 1-RDM and natural orbitals as a function of time; the equations of motion for these two quantities are given by Eqs. (26) and (21) in the main text, respectively. 
To solve for these 5 coupled equations, it is also necessary to solve, at each point in time, for the embedding Hamiltonian along with the operators $\hat{X}$, $\hat{G}$, and the generalized Fock matrix, $F$, given by Eqs. (20) and (23) in the main text and Eqs. (32)-(34) in the SI, respectively. This procedure can be summarized as follows:\\
All equations referenced in the following algorithm are contained within the SI unless specified otherwise.
\begin{enumerate}

    \item Given $C(t)$, $R(t)$, $\rho^{mf}(t)$, $\rho^{glob}(t)$, and $U(t)$
    \item Use $R(t)$ to generate the embedding Hamiltonian.\cite{wouters_practical_2016}
    \item Use $C(t)$ and the embedding Hamiltonian to calculate $F(t)$ using Eqs. \eqref{eq:SI_mod_F}-\eqref{eq:SI_act_F}
    \item Use $F(t)$ and $R(t)$ to calculate $\dot{\rho}^{glob}(t)$ using Eq. \eqref{eq:SI_td_rho_glob} 
    \item Use $U(t)$ and $\rho^{glob}(t)$ to obtain the eigenvalues of $\rho^{glob}(t)$ using Eq. \eqref{eq:SI_RDM_glob_and_U}
    \item Use $U(t)$, $\dot{\rho}^{glob}(t)$, and the eigenvalues from step 5 to obtain $G(t)$ using Eq. \eqref{eq:SI_G_def}
    \item Use $G(t)$ to obtain $\dot{U}(t)$ using Eq. \eqref{eq:SI_td_U}
    \item Use $\rho^{mf}(t)$ and $G(t)$ to obtain $\dot{\rho}^{mf}(t)$ using Eq. \eqref{eq:SI_td_rho_mf_final}
    \item Use $R(t)$ and $\rho^{mf}(t)$ to obtain the eigenvalues of $\rho^{mf}(t)$ using Eq. \eqref{eq:SI_RDM_mf_and_R} 
    \item Use $R(t)$, $\dot{\rho}^{mf}(t)$, and the eigenvalues from step 9 to obtain $X(t)$ using Eq. \eqref{eq:SI_td_lambda_with_X_and_RDM_final}
    \item Use $X(t)$ to obtain $\dot{R}$(t) using Eq. \eqref{eq:SI_td_R} 
    \item Use $X(t)$ and the embedding Hamiltonian to obtain $\dot{C}$(t) using Eq. (19) from the main text
\end{enumerate}
In principle, any possible numerical integrator could be used to solve for the equations of motion once the desired time-derivatives are calculated following the procedure above. In the context of this work, the 5 coupled equations of motions are numerically solved using the fourth order Runge-Kutta method. This implies that the above procedure is performed at each step within the Runge-Kutta routine.

While theoretically it is only necessary to propagate $\rho^{mf}$ and $C$, propagation of the embedding and natural orbitals is implemented to enhance the numerical stability of the method in a highly degenerate space. All calculations are initialized from the ground state where Coulomb interactions are absent (U=0). The solution of the static pDMET is exact for this state which eliminates any potential discrepancy between methods due to the error in initial state calculation; this choice allows for a direct comparison of the accuracy of solely the dynamics. However, this initialization procedure means that many of the eigenvectors of $\rho^{mf}$ and $\rho^{global}$ are degenerate. While both embedding and natural orbitals can be generated directly by diagonalizing previously mentioned matrices, it becomes numerically challenging to obtain smoothly transitioning eigenvectors at each time step when eigenvectors are nearly degenerate. It was observed that direct propagation of $\rho^{global}$ helps to ensure that the condition stated in Eq. (9) in the main text is maintained throughout the calculation. 

The time-dependence of nearly-degenerate embedding and natural orbitals introduces additional numerically challenges associated with the definition of corresponding operators $\hat{X}$ and $\hat{G}$ (Eq. \eqref{eq:SI_td_lambda_with_X_and_RDM_final}, \eqref{eq:SI_G_def} in SI ). Both matrices become undefined when orbitals are strictly degenerate. This gauge invariance is fixed by setting the corresponding elements of the $X$ or $G$ matrices to 0. However, direct treatment of near-degenerate eigenvectors results in numerically unstable terms and hence rapid divergence between previously maintained matching conditions. To address this, regularization parameters $\varepsilon_X$ and $\varepsilon_G$ are introduced for each matrix, such that $X^A_{ba}=0$ and $G_{\mu\nu}=0$ if $|\lambda_a-\lambda_b|<\varepsilon_X$ and $|\lambda_\mu-\lambda_\nu|<\varepsilon_G$, respectively.
Both parameters define a near-degenerate region around strictly equivalent eigenvalues where the elements are still set to 0. 
Similar parameters have been utilized in the context of TD-CASSCF and MCTDH.\cite{kretchmer_real-time_2018,Kre18b,manthe_wavepacket_1992,meyer_multi-configurational_1990} 
It was observed that only a small value of $\varepsilon_X$ was necessary to obtain numerically stable dynamics. By contrast, the numerical stability was more sensitive to the choice of $\varepsilon_G$ and a larger value was necessary. This, however, may be an artifact due to the high-degeneracy of the mean-field description of the system at $t=0$ observed in the systems studied in this work; all calculations are initialized from non-interacting ground-states, such that the natural orbitals all start off to be either completely occupied or unoccupied. 

\subsection{Simulation parameters}

The static pDMET calculations were initialized with a mean-field 1-RDM obtained from a Restricted Hartree Fock calculation. Real-time pDMET was performed using a time step of dt=0.0001 for the majority of calculations. However, the calculations with strong correlation effects ($U'=3,4$) and a large fragment size ($N_{frag}=5$) required an adaptive time step, where dt was reduced to dt=0.00001 when any two eigenvalues of the natural orbitals are within $2*10^{-5}$.
The stabilizing parameter for G matrix was set to $\varepsilon_{G}=10^{-5}$ and the analogous parameter for X matrix was $\varepsilon_{X}=10^{-9}$. 

TD-DMRG calculations were initialized and propagated in time using the ITensor package \cite{itensor}. Initial static DMRG calculations for the SIAM of N=60 sites and the MIAM of N=120 sites were done using 6 sweeps and a truncation error of $10^{-10}$. The bond dimension was progressively increased during the 6 sweeps using the following progression: 10, 20, 100, 100, 200, 500. The TD-DMRG calculation was performed using direct time-evolution of the matrix product state using an exponentiated matrix product operator.

All calculations were treated with the ``Fit" method within ITensor, where the exponential propagator is approximated to be equal $e^{-itH}$ up to terms on the order of $t^2$. The corresponding parameters are ``Cutoff"= $10^{-7}$, ``MaxDim" = 500, and dt=0.001. Convergence with respect to the cutoff, time step, and bond dimension was observed. 

Time-dependent Hartree Fock calculations are initialized from a Restricted Hartree Fock calculation to form the initial 1-RDM. The 1-RDM is propagated using a time-dependent Fock operator using the fourth-order Runge-Kutta method. All calculations were performed with a time step of dt=0.001. 

\section{References}
\bibliographystyle{apsrev4-1}
\bibliography{combined_biblio_final}

%% file: main_combined.bbl
\begin{thebibliography}{86}%
\makeatletter
\providecommand \@ifxundefined [1]{%
 \@ifx{#1\undefined}
}%
\providecommand \@ifnum [1]{%
 \ifnum #1\expandafter \@firstoftwo
 \else \expandafter \@secondoftwo
 \fi
}%
\providecommand \@ifx [1]{%
 \ifx #1\expandafter \@firstoftwo
 \else \expandafter \@secondoftwo
 \fi
}%
\providecommand \natexlab [1]{#1}%
\providecommand \enquote  [1]{``#1''}%
\providecommand \bibnamefont  [1]{#1}%
\providecommand \bibfnamefont [1]{#1}%
\providecommand \citenamefont [1]{#1}%
\providecommand \href@noop [0]{\@secondoftwo}%
\providecommand \href [0]{\begingroup \@sanitize@url \@href}%
\providecommand \@href[1]{\@@startlink{#1}\@@href}%
\providecommand \@@href[1]{\endgroup#1\@@endlink}%
\providecommand \@sanitize@url [0]{\catcode `\\12\catcode `\$12\catcode
  `\&12\catcode `\#12\catcode `\^12\catcode `\_12\catcode `\%12\relax}%
\providecommand \@@startlink[1]{}%
\providecommand \@@endlink[0]{}%
\providecommand \url  [0]{\begingroup\@sanitize@url \@url }%
\providecommand \@url [1]{\endgroup\@href {#1}{\urlprefix }}%
\providecommand \urlprefix  [0]{URL }%
\providecommand \Eprint [0]{\href }%
\providecommand \doibase [0]{http://dx.doi.org/}%
\providecommand \selectlanguage [0]{\@gobble}%
\providecommand \bibinfo  [0]{\@secondoftwo}%
\providecommand \bibfield  [0]{\@secondoftwo}%
\providecommand \translation [1]{[#1]}%
\providecommand \BibitemOpen [0]{}%
\providecommand \bibitemStop [0]{}%
\providecommand \bibitemNoStop [0]{.\EOS\space}%
\providecommand \EOS [0]{\spacefactor3000\relax}%
\providecommand \BibitemShut  [1]{\csname bibitem#1\endcsname}%
\let\auto@bib@innerbib\@empty
\bibitem [{\citenamefont {Peng}\ \emph {et~al.}(2019)\citenamefont {Peng},
  \citenamefont {Marceau},\ and\ \citenamefont {Villeneuve}}]{Pen19}%
  \BibitemOpen
  \bibfield  {author} {\bibinfo {author} {\bibfnamefont {P.}~\bibnamefont
  {Peng}}, \bibinfo {author} {\bibfnamefont {C.}~\bibnamefont {Marceau}}, \
  and\ \bibinfo {author} {\bibfnamefont {D.~M.}\ \bibnamefont {Villeneuve}},\
  }\href@noop {} {\bibfield  {journal} {\bibinfo  {journal} {Nature Reviews
  Physics}\ }\textbf {\bibinfo {volume} {1}},\ \bibinfo {pages} {144} (\bibinfo
  {year} {2019})}\BibitemShut {NoStop}%
\bibitem [{\citenamefont {Nisoli}\ \emph {et~al.}(2017)\citenamefont {Nisoli},
  \citenamefont {Decleva}, \citenamefont {Calegari}, \citenamefont {Palacios},\
  and\ \citenamefont {Mart{\'\i}n}}]{nis17}%
  \BibitemOpen
  \bibfield  {author} {\bibinfo {author} {\bibfnamefont {M.}~\bibnamefont
  {Nisoli}}, \bibinfo {author} {\bibfnamefont {P.}~\bibnamefont {Decleva}},
  \bibinfo {author} {\bibfnamefont {F.}~\bibnamefont {Calegari}}, \bibinfo
  {author} {\bibfnamefont {A.}~\bibnamefont {Palacios}}, \ and\ \bibinfo
  {author} {\bibfnamefont {F.}~\bibnamefont {Mart{\'\i}n}},\ }\href@noop {}
  {\bibfield  {journal} {\bibinfo  {journal} {Chemical reviews}\ }\textbf
  {\bibinfo {volume} {117}},\ \bibinfo {pages} {10760} (\bibinfo {year}
  {2017})}\BibitemShut {NoStop}%
\bibitem [{\citenamefont {Varillas}\ \emph {et~al.}(2022)\citenamefont
  {Varillas}, \citenamefont {Lucchini},\ and\ \citenamefont {Nisoli}}]{Var22}%
  \BibitemOpen
  \bibfield  {author} {\bibinfo {author} {\bibfnamefont {R.~B.}\ \bibnamefont
  {Varillas}}, \bibinfo {author} {\bibfnamefont {M.}~\bibnamefont {Lucchini}},
  \ and\ \bibinfo {author} {\bibfnamefont {M.}~\bibnamefont {Nisoli}},\
  }\href@noop {} {\bibfield  {journal} {\bibinfo  {journal} {Reports on
  Progress in Physics}\ } (\bibinfo {year} {2022})}\BibitemShut {NoStop}%
\bibitem [{\citenamefont {Gallmann}\ \emph {et~al.}(2012)\citenamefont
  {Gallmann}, \citenamefont {Cirelli},\ and\ \citenamefont {Keller}}]{Gal12}%
  \BibitemOpen
  \bibfield  {author} {\bibinfo {author} {\bibfnamefont {L.}~\bibnamefont
  {Gallmann}}, \bibinfo {author} {\bibfnamefont {C.}~\bibnamefont {Cirelli}}, \
  and\ \bibinfo {author} {\bibfnamefont {U.}~\bibnamefont {Keller}},\
  }\href@noop {} {\bibfield  {journal} {\bibinfo  {journal} {Annual review of
  physical chemistry}\ }\textbf {\bibinfo {volume} {63}},\ \bibinfo {pages}
  {447} (\bibinfo {year} {2012})}\BibitemShut {NoStop}%
\bibitem [{\citenamefont {M{\aa}nsson}\ \emph {et~al.}(2021)\citenamefont
  {M{\aa}nsson}, \citenamefont {Latini}, \citenamefont {Covito}, \citenamefont
  {Wanie}, \citenamefont {Galli}, \citenamefont {Perfetto}, \citenamefont
  {Stefanucci}, \citenamefont {H{\"u}bener}, \citenamefont {De~Giovannini},
  \citenamefont {Castrovilli} \emph {et~al.}}]{Maa21}%
  \BibitemOpen
  \bibfield  {author} {\bibinfo {author} {\bibfnamefont {E.~P.}\ \bibnamefont
  {M{\aa}nsson}}, \bibinfo {author} {\bibfnamefont {S.}~\bibnamefont {Latini}},
  \bibinfo {author} {\bibfnamefont {F.}~\bibnamefont {Covito}}, \bibinfo
  {author} {\bibfnamefont {V.}~\bibnamefont {Wanie}}, \bibinfo {author}
  {\bibfnamefont {M.}~\bibnamefont {Galli}}, \bibinfo {author} {\bibfnamefont
  {E.}~\bibnamefont {Perfetto}}, \bibinfo {author} {\bibfnamefont
  {G.}~\bibnamefont {Stefanucci}}, \bibinfo {author} {\bibfnamefont
  {H.}~\bibnamefont {H{\"u}bener}}, \bibinfo {author} {\bibfnamefont
  {U.}~\bibnamefont {De~Giovannini}}, \bibinfo {author} {\bibfnamefont {M.~C.}\
  \bibnamefont {Castrovilli}},  \emph {et~al.},\ }\href@noop {} {\bibfield
  {journal} {\bibinfo  {journal} {Communications Chemistry}\ }\textbf {\bibinfo
  {volume} {4}},\ \bibinfo {pages} {1} (\bibinfo {year} {2021})}\BibitemShut
  {NoStop}%
\bibitem [{\citenamefont {de~la Cruz~Valbuena}\ \emph
  {et~al.}(2019)\citenamefont {de~la Cruz~Valbuena}, \citenamefont
  {VA~Camargo}, \citenamefont {Borrego-Varillas}, \citenamefont {Perozeni},
  \citenamefont {D’Andrea}, \citenamefont {Ballottari},\ and\ \citenamefont
  {Cerullo}}]{De19}%
  \BibitemOpen
  \bibfield  {author} {\bibinfo {author} {\bibfnamefont {G.}~\bibnamefont
  {de~la Cruz~Valbuena}}, \bibinfo {author} {\bibfnamefont {F.}~\bibnamefont
  {VA~Camargo}}, \bibinfo {author} {\bibfnamefont {R.}~\bibnamefont
  {Borrego-Varillas}}, \bibinfo {author} {\bibfnamefont {F.}~\bibnamefont
  {Perozeni}}, \bibinfo {author} {\bibfnamefont {C.}~\bibnamefont
  {D’Andrea}}, \bibinfo {author} {\bibfnamefont {M.}~\bibnamefont
  {Ballottari}}, \ and\ \bibinfo {author} {\bibfnamefont {G.}~\bibnamefont
  {Cerullo}},\ }\href@noop {} {\bibfield  {journal} {\bibinfo  {journal} {The
  journal of physical chemistry letters}\ }\textbf {\bibinfo {volume} {10}},\
  \bibinfo {pages} {2500} (\bibinfo {year} {2019})}\BibitemShut {NoStop}%
\bibitem [{\citenamefont {Krueger}\ \emph {et~al.}(2020)\citenamefont
  {Krueger}, \citenamefont {Boulanger}, \citenamefont {Zhu}, \citenamefont
  {Tang},\ and\ \citenamefont {Fang}}]{Kru20}%
  \BibitemOpen
  \bibfield  {author} {\bibinfo {author} {\bibfnamefont {T.~D.}\ \bibnamefont
  {Krueger}}, \bibinfo {author} {\bibfnamefont {S.~A.}\ \bibnamefont
  {Boulanger}}, \bibinfo {author} {\bibfnamefont {L.}~\bibnamefont {Zhu}},
  \bibinfo {author} {\bibfnamefont {L.}~\bibnamefont {Tang}}, \ and\ \bibinfo
  {author} {\bibfnamefont {C.}~\bibnamefont {Fang}},\ }\href@noop {} {\bibfield
   {journal} {\bibinfo  {journal} {Structural Dynamics}\ }\textbf {\bibinfo
  {volume} {7}},\ \bibinfo {pages} {024901} (\bibinfo {year}
  {2020})}\BibitemShut {NoStop}%
\bibitem [{\citenamefont {Fuemmeler}\ \emph {et~al.}(2016)\citenamefont
  {Fuemmeler}, \citenamefont {Sanders}, \citenamefont {Pun}, \citenamefont
  {Kumarasamy}, \citenamefont {Zeng}, \citenamefont {Miyata}, \citenamefont
  {Steigerwald}, \citenamefont {Zhu}, \citenamefont {Sfeir}, \citenamefont
  {Campos} \emph {et~al.}}]{Fue16}%
  \BibitemOpen
  \bibfield  {author} {\bibinfo {author} {\bibfnamefont {E.~G.}\ \bibnamefont
  {Fuemmeler}}, \bibinfo {author} {\bibfnamefont {S.~N.}\ \bibnamefont
  {Sanders}}, \bibinfo {author} {\bibfnamefont {A.~B.}\ \bibnamefont {Pun}},
  \bibinfo {author} {\bibfnamefont {E.}~\bibnamefont {Kumarasamy}}, \bibinfo
  {author} {\bibfnamefont {T.}~\bibnamefont {Zeng}}, \bibinfo {author}
  {\bibfnamefont {K.}~\bibnamefont {Miyata}}, \bibinfo {author} {\bibfnamefont
  {M.~L.}\ \bibnamefont {Steigerwald}}, \bibinfo {author} {\bibfnamefont
  {X.-Y.}\ \bibnamefont {Zhu}}, \bibinfo {author} {\bibfnamefont {M.~Y.}\
  \bibnamefont {Sfeir}}, \bibinfo {author} {\bibfnamefont {L.~M.}\ \bibnamefont
  {Campos}},  \emph {et~al.},\ }\href@noop {} {\bibfield  {journal} {\bibinfo
  {journal} {ACS central science}\ }\textbf {\bibinfo {volume} {2}},\ \bibinfo
  {pages} {316} (\bibinfo {year} {2016})}\BibitemShut {NoStop}%
\bibitem [{\citenamefont {Ponseca}\ \emph {et~al.}(2017)\citenamefont
  {Ponseca}, \citenamefont {Chábera}, \citenamefont {Uhlig}, \citenamefont
  {Persson},\ and\ \citenamefont {Sundström}}]{ponseca_ultrafast_2017}%
  \BibitemOpen
  \bibfield  {author} {\bibinfo {author} {\bibfnamefont {C.~S.}\ \bibnamefont
  {Ponseca}}, \bibinfo {author} {\bibfnamefont {P.}~\bibnamefont {Chábera}},
  \bibinfo {author} {\bibfnamefont {J.}~\bibnamefont {Uhlig}}, \bibinfo
  {author} {\bibfnamefont {P.}~\bibnamefont {Persson}}, \ and\ \bibinfo
  {author} {\bibfnamefont {V.}~\bibnamefont {Sundström}},\ }\href {\doibase
  10.1021/acs.chemrev.6b00807} {\bibfield  {journal} {\bibinfo  {journal}
  {Chemical Reviews}\ }\textbf {\bibinfo {volume} {117}},\ \bibinfo {pages}
  {10940} (\bibinfo {year} {2017})}\BibitemShut {NoStop}%
\bibitem [{\citenamefont {Cui}\ \emph {et~al.}(2019)\citenamefont {Cui},
  \citenamefont {Hur}, \citenamefont {Akbar}, \citenamefont {Klöckner},
  \citenamefont {Jeong}, \citenamefont {Pauly}, \citenamefont {Jang},
  \citenamefont {Reddy},\ and\ \citenamefont {Meyhofer}}]{cui_thermal_2019}%
  \BibitemOpen
  \bibfield  {author} {\bibinfo {author} {\bibfnamefont {L.}~\bibnamefont
  {Cui}}, \bibinfo {author} {\bibfnamefont {S.}~\bibnamefont {Hur}}, \bibinfo
  {author} {\bibfnamefont {Z.~A.}\ \bibnamefont {Akbar}}, \bibinfo {author}
  {\bibfnamefont {J.~C.}\ \bibnamefont {Klöckner}}, \bibinfo {author}
  {\bibfnamefont {W.}~\bibnamefont {Jeong}}, \bibinfo {author} {\bibfnamefont
  {F.}~\bibnamefont {Pauly}}, \bibinfo {author} {\bibfnamefont {S.-Y.}\
  \bibnamefont {Jang}}, \bibinfo {author} {\bibfnamefont {P.}~\bibnamefont
  {Reddy}}, \ and\ \bibinfo {author} {\bibfnamefont {E.}~\bibnamefont
  {Meyhofer}},\ }\href {\doibase 10.1038/s41586-019-1420-z} {\bibfield
  {journal} {\bibinfo  {journal} {Nature}\ }\textbf {\bibinfo {volume} {572}},\
  \bibinfo {pages} {628} (\bibinfo {year} {2019})},\ \bibinfo {note} {number:
  7771 Publisher: Nature Publishing Group}\BibitemShut {NoStop}%
\bibitem [{\citenamefont {Joachim}\ \emph {et~al.}(2000)\citenamefont
  {Joachim}, \citenamefont {Gimzewski},\ and\ \citenamefont
  {Aviram}}]{joachim_electronics_2000}%
  \BibitemOpen
  \bibfield  {author} {\bibinfo {author} {\bibfnamefont {C.}~\bibnamefont
  {Joachim}}, \bibinfo {author} {\bibfnamefont {J.~K.}\ \bibnamefont
  {Gimzewski}}, \ and\ \bibinfo {author} {\bibfnamefont {A.}~\bibnamefont
  {Aviram}},\ }\href {\doibase 10.1038/35046000} {\bibfield  {journal}
  {\bibinfo  {journal} {Nature}\ }\textbf {\bibinfo {volume} {408}},\ \bibinfo
  {pages} {541} (\bibinfo {year} {2000})},\ \bibinfo {note} {number: 6812
  Publisher: Nature Publishing Group}\BibitemShut {NoStop}%
\bibitem [{\citenamefont {Nitzan}\ and\ \citenamefont
  {Ratner}(2003)}]{nitzan_electron_2003}%
  \BibitemOpen
  \bibfield  {author} {\bibinfo {author} {\bibfnamefont {A.}~\bibnamefont
  {Nitzan}}\ and\ \bibinfo {author} {\bibfnamefont {M.~A.}\ \bibnamefont
  {Ratner}},\ }\href {\doibase 10.1126/science.1081572} {\bibfield  {journal}
  {\bibinfo  {journal} {Science}\ }\textbf {\bibinfo {volume} {300}},\ \bibinfo
  {pages} {1384} (\bibinfo {year} {2003})}\BibitemShut {NoStop}%
\bibitem [{\citenamefont {Zhang}\ \emph {et~al.}(2021)\citenamefont {Zhang},
  \citenamefont {Zhou}, \citenamefont {Li}, \citenamefont {Shen},\ and\
  \citenamefont {Feng}}]{zhang_recent_2021}%
  \BibitemOpen
  \bibfield  {author} {\bibinfo {author} {\bibfnamefont {L.}~\bibnamefont
  {Zhang}}, \bibinfo {author} {\bibfnamefont {J.}~\bibnamefont {Zhou}},
  \bibinfo {author} {\bibfnamefont {H.}~\bibnamefont {Li}}, \bibinfo {author}
  {\bibfnamefont {L.}~\bibnamefont {Shen}}, \ and\ \bibinfo {author}
  {\bibfnamefont {Y.~P.}\ \bibnamefont {Feng}},\ }\href {\doibase
  10.1063/5.0032538} {\bibfield  {journal} {\bibinfo  {journal} {Applied
  Physics Reviews}\ }\textbf {\bibinfo {volume} {8}},\ \bibinfo {pages}
  {021308} (\bibinfo {year} {2021})}\BibitemShut {NoStop}%
\bibitem [{\citenamefont {Dayen}\ \emph {et~al.}(2020)\citenamefont {Dayen},
  \citenamefont {Ray}, \citenamefont {Karis}, \citenamefont {Vera-Marun},\ and\
  \citenamefont {Kamalakar}}]{dayen_two-dimensional_2020}%
  \BibitemOpen
  \bibfield  {author} {\bibinfo {author} {\bibfnamefont {J.-F.}\ \bibnamefont
  {Dayen}}, \bibinfo {author} {\bibfnamefont {S.~J.}\ \bibnamefont {Ray}},
  \bibinfo {author} {\bibfnamefont {O.}~\bibnamefont {Karis}}, \bibinfo
  {author} {\bibfnamefont {I.~J.}\ \bibnamefont {Vera-Marun}}, \ and\ \bibinfo
  {author} {\bibfnamefont {M.~V.}\ \bibnamefont {Kamalakar}},\ }\href {\doibase
  10.1063/1.5112171} {\bibfield  {journal} {\bibinfo  {journal} {Applied
  Physics Reviews}\ }\textbf {\bibinfo {volume} {7}},\ \bibinfo {pages}
  {011303} (\bibinfo {year} {2020})}\BibitemShut {NoStop}%
\bibitem [{\citenamefont {Singh}\ \emph {et~al.}(2019)\citenamefont {Singh},
  \citenamefont {Liu}, \citenamefont {Lim}, \citenamefont {Robel},\ and\
  \citenamefont {Klimov}}]{singh_hot-electron_2019}%
  \BibitemOpen
  \bibfield  {author} {\bibinfo {author} {\bibfnamefont {R.}~\bibnamefont
  {Singh}}, \bibinfo {author} {\bibfnamefont {W.}~\bibnamefont {Liu}}, \bibinfo
  {author} {\bibfnamefont {J.}~\bibnamefont {Lim}}, \bibinfo {author}
  {\bibfnamefont {I.}~\bibnamefont {Robel}}, \ and\ \bibinfo {author}
  {\bibfnamefont {V.~I.}\ \bibnamefont {Klimov}},\ }\href {\doibase
  10.1038/s41565-019-0548-1} {\bibfield  {journal} {\bibinfo  {journal} {Nature
  Nanotechnology}\ }\textbf {\bibinfo {volume} {14}},\ \bibinfo {pages} {1035}
  (\bibinfo {year} {2019})},\ \bibinfo {note} {number: 11 Publisher: Nature
  Publishing Group}\BibitemShut {NoStop}%
\bibitem [{\citenamefont {Bande}(2013)}]{bande_electron_2013}%
  \BibitemOpen
  \bibfield  {author} {\bibinfo {author} {\bibfnamefont {A.}~\bibnamefont
  {Bande}},\ }\href {\doibase 10.1063/1.4807611} {\bibfield  {journal}
  {\bibinfo  {journal} {The Journal of Chemical Physics}\ }\textbf {\bibinfo
  {volume} {138}},\ \bibinfo {pages} {214104} (\bibinfo {year}
  {2013})}\BibitemShut {NoStop}%
\bibitem [{\citenamefont {Jahnke}\ \emph {et~al.}(2020)\citenamefont {Jahnke},
  \citenamefont {Hergenhahn}, \citenamefont {Winter}, \citenamefont {Dörner},
  \citenamefont {Frühling}, \citenamefont {Demekhin}, \citenamefont
  {Gokhberg}, \citenamefont {Cederbaum}, \citenamefont {Ehresmann},
  \citenamefont {Knie},\ and\ \citenamefont {Dreuw}}]{jahnke_interatomic_2020}%
  \BibitemOpen
  \bibfield  {author} {\bibinfo {author} {\bibfnamefont {T.}~\bibnamefont
  {Jahnke}}, \bibinfo {author} {\bibfnamefont {U.}~\bibnamefont {Hergenhahn}},
  \bibinfo {author} {\bibfnamefont {B.}~\bibnamefont {Winter}}, \bibinfo
  {author} {\bibfnamefont {R.}~\bibnamefont {Dörner}}, \bibinfo {author}
  {\bibfnamefont {U.}~\bibnamefont {Frühling}}, \bibinfo {author}
  {\bibfnamefont {P.~V.}\ \bibnamefont {Demekhin}}, \bibinfo {author}
  {\bibfnamefont {K.}~\bibnamefont {Gokhberg}}, \bibinfo {author}
  {\bibfnamefont {L.~S.}\ \bibnamefont {Cederbaum}}, \bibinfo {author}
  {\bibfnamefont {A.}~\bibnamefont {Ehresmann}}, \bibinfo {author}
  {\bibfnamefont {A.}~\bibnamefont {Knie}}, \ and\ \bibinfo {author}
  {\bibfnamefont {A.}~\bibnamefont {Dreuw}},\ }\href {\doibase
  10.1021/acs.chemrev.0c00106} {\bibfield  {journal} {\bibinfo  {journal}
  {Chemical Reviews}\ }\textbf {\bibinfo {volume} {120}},\ \bibinfo {pages}
  {11295} (\bibinfo {year} {2020})}\BibitemShut {NoStop}%
\bibitem [{\citenamefont {Demekhin}\ \emph {et~al.}(2009)\citenamefont
  {Demekhin}, \citenamefont {Chiang}, \citenamefont {Stoychev}, \citenamefont
  {Kolorenč}, \citenamefont {Scheit}, \citenamefont {Kuleff}, \citenamefont
  {Tarantelli},\ and\ \citenamefont {Cederbaum}}]{demekhin_interatomic_2009}%
  \BibitemOpen
  \bibfield  {author} {\bibinfo {author} {\bibfnamefont {P.~V.}\ \bibnamefont
  {Demekhin}}, \bibinfo {author} {\bibfnamefont {Y.-C.}\ \bibnamefont
  {Chiang}}, \bibinfo {author} {\bibfnamefont {S.~D.}\ \bibnamefont
  {Stoychev}}, \bibinfo {author} {\bibfnamefont {P.}~\bibnamefont {Kolorenč}},
  \bibinfo {author} {\bibfnamefont {S.}~\bibnamefont {Scheit}}, \bibinfo
  {author} {\bibfnamefont {A.~I.}\ \bibnamefont {Kuleff}}, \bibinfo {author}
  {\bibfnamefont {F.}~\bibnamefont {Tarantelli}}, \ and\ \bibinfo {author}
  {\bibfnamefont {L.~S.}\ \bibnamefont {Cederbaum}},\ }\href {\doibase
  10.1063/1.3211114} {\bibfield  {journal} {\bibinfo  {journal} {The Journal of
  Chemical Physics}\ }\textbf {\bibinfo {volume} {131}},\ \bibinfo {pages}
  {104303} (\bibinfo {year} {2009})}\BibitemShut {NoStop}%
\bibitem [{\citenamefont {Ouchi}\ \emph {et~al.}(2011)\citenamefont {Ouchi},
  \citenamefont {Sakai}, \citenamefont {Fukuzawa}, \citenamefont {Higuchi},
  \citenamefont {Demekhin}, \citenamefont {Chiang}, \citenamefont {Stoychev},
  \citenamefont {Kuleff}, \citenamefont {Mazza}, \citenamefont {Schöffler},
  \citenamefont {Nagaya}, \citenamefont {Yao}, \citenamefont {Tamenori},
  \citenamefont {Saito},\ and\ \citenamefont {Ueda}}]{ouchi_interatomic_2011}%
  \BibitemOpen
  \bibfield  {author} {\bibinfo {author} {\bibfnamefont {T.}~\bibnamefont
  {Ouchi}}, \bibinfo {author} {\bibfnamefont {K.}~\bibnamefont {Sakai}},
  \bibinfo {author} {\bibfnamefont {H.}~\bibnamefont {Fukuzawa}}, \bibinfo
  {author} {\bibfnamefont {I.}~\bibnamefont {Higuchi}}, \bibinfo {author}
  {\bibfnamefont {P.~V.}\ \bibnamefont {Demekhin}}, \bibinfo {author}
  {\bibfnamefont {Y.-C.}\ \bibnamefont {Chiang}}, \bibinfo {author}
  {\bibfnamefont {S.~D.}\ \bibnamefont {Stoychev}}, \bibinfo {author}
  {\bibfnamefont {A.~I.}\ \bibnamefont {Kuleff}}, \bibinfo {author}
  {\bibfnamefont {T.}~\bibnamefont {Mazza}}, \bibinfo {author} {\bibfnamefont
  {M.}~\bibnamefont {Schöffler}}, \bibinfo {author} {\bibfnamefont
  {K.}~\bibnamefont {Nagaya}}, \bibinfo {author} {\bibfnamefont
  {M.}~\bibnamefont {Yao}}, \bibinfo {author} {\bibfnamefont {Y.}~\bibnamefont
  {Tamenori}}, \bibinfo {author} {\bibfnamefont {N.}~\bibnamefont {Saito}}, \
  and\ \bibinfo {author} {\bibfnamefont {K.}~\bibnamefont {Ueda}},\ }\href
  {\doibase 10.1103/PhysRevA.83.053415} {\bibfield  {journal} {\bibinfo
  {journal} {Physical Review A}\ }\textbf {\bibinfo {volume} {83}},\ \bibinfo
  {pages} {053415} (\bibinfo {year} {2011})}\BibitemShut {NoStop}%
\bibitem [{\citenamefont {Aoki}\ \emph {et~al.}(2014)\citenamefont {Aoki},
  \citenamefont {Tsuji}, \citenamefont {Eckstein}, \citenamefont {Kollar},
  \citenamefont {Oka},\ and\ \citenamefont {Werner}}]{Aok14}%
  \BibitemOpen
  \bibfield  {author} {\bibinfo {author} {\bibfnamefont {H.}~\bibnamefont
  {Aoki}}, \bibinfo {author} {\bibfnamefont {N.}~\bibnamefont {Tsuji}},
  \bibinfo {author} {\bibfnamefont {M.}~\bibnamefont {Eckstein}}, \bibinfo
  {author} {\bibfnamefont {M.}~\bibnamefont {Kollar}}, \bibinfo {author}
  {\bibfnamefont {T.}~\bibnamefont {Oka}}, \ and\ \bibinfo {author}
  {\bibfnamefont {P.}~\bibnamefont {Werner}},\ }\href@noop {} {\bibfield
  {journal} {\bibinfo  {journal} {Reviews of Modern Physics}\ }\textbf
  {\bibinfo {volume} {86}},\ \bibinfo {pages} {779} (\bibinfo {year}
  {2014})}\BibitemShut {NoStop}%
\bibitem [{\citenamefont {H{\"a}rtle}\ \emph {et~al.}(2008)\citenamefont
  {H{\"a}rtle}, \citenamefont {Benesch},\ and\ \citenamefont {Thoss}}]{Har08}%
  \BibitemOpen
  \bibfield  {author} {\bibinfo {author} {\bibfnamefont {R.}~\bibnamefont
  {H{\"a}rtle}}, \bibinfo {author} {\bibfnamefont {C.}~\bibnamefont {Benesch}},
  \ and\ \bibinfo {author} {\bibfnamefont {M.}~\bibnamefont {Thoss}},\
  }\href@noop {} {\bibfield  {journal} {\bibinfo  {journal} {Physical Review
  B}\ }\textbf {\bibinfo {volume} {77}},\ \bibinfo {pages} {205314} (\bibinfo
  {year} {2008})}\BibitemShut {NoStop}%
\bibitem [{\citenamefont {Rammer}(2018)}]{Ram18}%
  \BibitemOpen
  \bibfield  {author} {\bibinfo {author} {\bibfnamefont {J.}~\bibnamefont
  {Rammer}},\ }\href@noop {} {\emph {\bibinfo {title} {Quantum transport
  theory}}}\ (\bibinfo  {publisher} {CRC Press},\ \bibinfo {year}
  {2018})\BibitemShut {NoStop}%
\bibitem [{\citenamefont {Freericks}\ \emph {et~al.}(2006)\citenamefont
  {Freericks}, \citenamefont {Turkowski},\ and\ \citenamefont
  {Zlati{\'c}}}]{Fre06}%
  \BibitemOpen
  \bibfield  {author} {\bibinfo {author} {\bibfnamefont {J.}~\bibnamefont
  {Freericks}}, \bibinfo {author} {\bibfnamefont {V.}~\bibnamefont
  {Turkowski}}, \ and\ \bibinfo {author} {\bibfnamefont {V.}~\bibnamefont
  {Zlati{\'c}}},\ }\href@noop {} {\bibfield  {journal} {\bibinfo  {journal}
  {Physical review letters}\ }\textbf {\bibinfo {volume} {97}},\ \bibinfo
  {pages} {266408} (\bibinfo {year} {2006})}\BibitemShut {NoStop}%
\bibitem [{\citenamefont {Li}\ \emph {et~al.}(2013)\citenamefont {Li},
  \citenamefont {Levy}, \citenamefont {Swenson}, \citenamefont {Rabani},\ and\
  \citenamefont {Miller}}]{Li13}%
  \BibitemOpen
  \bibfield  {author} {\bibinfo {author} {\bibfnamefont {B.}~\bibnamefont
  {Li}}, \bibinfo {author} {\bibfnamefont {T.~J.}\ \bibnamefont {Levy}},
  \bibinfo {author} {\bibfnamefont {D.~W.}\ \bibnamefont {Swenson}}, \bibinfo
  {author} {\bibfnamefont {E.}~\bibnamefont {Rabani}}, \ and\ \bibinfo {author}
  {\bibfnamefont {W.~H.}\ \bibnamefont {Miller}},\ }\href@noop {} {\bibfield
  {journal} {\bibinfo  {journal} {The Journal of Chemical Physics}\ }\textbf
  {\bibinfo {volume} {138}},\ \bibinfo {pages} {104110} (\bibinfo {year}
  {2013})}\BibitemShut {NoStop}%
\bibitem [{\citenamefont {Swenson}\ \emph {et~al.}(2011)\citenamefont
  {Swenson}, \citenamefont {Levy}, \citenamefont {Cohen}, \citenamefont
  {Rabani},\ and\ \citenamefont {Miller}}]{Swe11}%
  \BibitemOpen
  \bibfield  {author} {\bibinfo {author} {\bibfnamefont {D.~W.}\ \bibnamefont
  {Swenson}}, \bibinfo {author} {\bibfnamefont {T.}~\bibnamefont {Levy}},
  \bibinfo {author} {\bibfnamefont {G.}~\bibnamefont {Cohen}}, \bibinfo
  {author} {\bibfnamefont {E.}~\bibnamefont {Rabani}}, \ and\ \bibinfo {author}
  {\bibfnamefont {W.~H.}\ \bibnamefont {Miller}},\ }\href@noop {} {\bibfield
  {journal} {\bibinfo  {journal} {The Journal of chemical physics}\ }\textbf
  {\bibinfo {volume} {134}},\ \bibinfo {pages} {164103} (\bibinfo {year}
  {2011})}\BibitemShut {NoStop}%
\bibitem [{\citenamefont {Swenson}\ \emph {et~al.}(2012)\citenamefont
  {Swenson}, \citenamefont {Cohen},\ and\ \citenamefont {Rabani}}]{Swe12}%
  \BibitemOpen
  \bibfield  {author} {\bibinfo {author} {\bibfnamefont {D.~W.}\ \bibnamefont
  {Swenson}}, \bibinfo {author} {\bibfnamefont {G.}~\bibnamefont {Cohen}}, \
  and\ \bibinfo {author} {\bibfnamefont {E.}~\bibnamefont {Rabani}},\
  }\href@noop {} {\bibfield  {journal} {\bibinfo  {journal} {Molecular
  Physics}\ }\textbf {\bibinfo {volume} {110}},\ \bibinfo {pages} {743}
  (\bibinfo {year} {2012})}\BibitemShut {NoStop}%
\bibitem [{\citenamefont {Cohen}\ \emph
  {et~al.}(2014{\natexlab{a}})\citenamefont {Cohen}, \citenamefont {Reichman},
  \citenamefont {Millis},\ and\ \citenamefont {Gull}}]{Coh14a}%
  \BibitemOpen
  \bibfield  {author} {\bibinfo {author} {\bibfnamefont {G.}~\bibnamefont
  {Cohen}}, \bibinfo {author} {\bibfnamefont {D.~R.}\ \bibnamefont {Reichman}},
  \bibinfo {author} {\bibfnamefont {A.~J.}\ \bibnamefont {Millis}}, \ and\
  \bibinfo {author} {\bibfnamefont {E.}~\bibnamefont {Gull}},\ }\href@noop {}
  {\bibfield  {journal} {\bibinfo  {journal} {Physical Review B}\ }\textbf
  {\bibinfo {volume} {89}},\ \bibinfo {pages} {115139} (\bibinfo {year}
  {2014}{\natexlab{a}})}\BibitemShut {NoStop}%
\bibitem [{\citenamefont {Cohen}\ \emph
  {et~al.}(2014{\natexlab{b}})\citenamefont {Cohen}, \citenamefont {Gull},
  \citenamefont {Reichman},\ and\ \citenamefont {Millis}}]{Coh14b}%
  \BibitemOpen
  \bibfield  {author} {\bibinfo {author} {\bibfnamefont {G.}~\bibnamefont
  {Cohen}}, \bibinfo {author} {\bibfnamefont {E.}~\bibnamefont {Gull}},
  \bibinfo {author} {\bibfnamefont {D.~R.}\ \bibnamefont {Reichman}}, \ and\
  \bibinfo {author} {\bibfnamefont {A.~J.}\ \bibnamefont {Millis}},\
  }\href@noop {} {\bibfield  {journal} {\bibinfo  {journal} {Physical review
  letters}\ }\textbf {\bibinfo {volume} {112}},\ \bibinfo {pages} {146802}
  (\bibinfo {year} {2014}{\natexlab{b}})}\BibitemShut {NoStop}%
\bibitem [{\citenamefont {Gull}\ \emph {et~al.}(2011)\citenamefont {Gull},
  \citenamefont {Millis}, \citenamefont {Lichtenstein}, \citenamefont
  {Rubtsov}, \citenamefont {Troyer},\ and\ \citenamefont {Werner}}]{Gul11}%
  \BibitemOpen
  \bibfield  {author} {\bibinfo {author} {\bibfnamefont {E.}~\bibnamefont
  {Gull}}, \bibinfo {author} {\bibfnamefont {A.~J.}\ \bibnamefont {Millis}},
  \bibinfo {author} {\bibfnamefont {A.~I.}\ \bibnamefont {Lichtenstein}},
  \bibinfo {author} {\bibfnamefont {A.~N.}\ \bibnamefont {Rubtsov}}, \bibinfo
  {author} {\bibfnamefont {M.}~\bibnamefont {Troyer}}, \ and\ \bibinfo {author}
  {\bibfnamefont {P.}~\bibnamefont {Werner}},\ }\href@noop {} {\bibfield
  {journal} {\bibinfo  {journal} {Reviews of Modern Physics}\ }\textbf
  {\bibinfo {volume} {83}},\ \bibinfo {pages} {349} (\bibinfo {year}
  {2011})}\BibitemShut {NoStop}%
\bibitem [{\citenamefont {Schir{\'o}}(2010)}]{Sch10a}%
  \BibitemOpen
  \bibfield  {author} {\bibinfo {author} {\bibfnamefont {M.}~\bibnamefont
  {Schir{\'o}}},\ }\href@noop {} {\bibfield  {journal} {\bibinfo  {journal}
  {Physical Review B}\ }\textbf {\bibinfo {volume} {81}},\ \bibinfo {pages}
  {085126} (\bibinfo {year} {2010})}\BibitemShut {NoStop}%
\bibitem [{\citenamefont {Werner}\ \emph {et~al.}(2009)\citenamefont {Werner},
  \citenamefont {Oka},\ and\ \citenamefont {Millis}}]{Wer09}%
  \BibitemOpen
  \bibfield  {author} {\bibinfo {author} {\bibfnamefont {P.}~\bibnamefont
  {Werner}}, \bibinfo {author} {\bibfnamefont {T.}~\bibnamefont {Oka}}, \ and\
  \bibinfo {author} {\bibfnamefont {A.~J.}\ \bibnamefont {Millis}},\
  }\href@noop {} {\bibfield  {journal} {\bibinfo  {journal} {Physical Review
  B}\ }\textbf {\bibinfo {volume} {79}},\ \bibinfo {pages} {035320} (\bibinfo
  {year} {2009})}\BibitemShut {NoStop}%
\bibitem [{\citenamefont {Cohen}\ \emph {et~al.}(2015)\citenamefont {Cohen},
  \citenamefont {Gull}, \citenamefont {Reichman},\ and\ \citenamefont
  {Millis}}]{Coh15}%
  \BibitemOpen
  \bibfield  {author} {\bibinfo {author} {\bibfnamefont {G.}~\bibnamefont
  {Cohen}}, \bibinfo {author} {\bibfnamefont {E.}~\bibnamefont {Gull}},
  \bibinfo {author} {\bibfnamefont {D.~R.}\ \bibnamefont {Reichman}}, \ and\
  \bibinfo {author} {\bibfnamefont {A.~J.}\ \bibnamefont {Millis}},\
  }\href@noop {} {\bibfield  {journal} {\bibinfo  {journal} {Physical review
  letters}\ }\textbf {\bibinfo {volume} {115}},\ \bibinfo {pages} {266802}
  (\bibinfo {year} {2015})}\BibitemShut {NoStop}%
\bibitem [{\citenamefont {M{\"u}hlbacher}\ and\ \citenamefont
  {Rabani}(2008)}]{Muh08}%
  \BibitemOpen
  \bibfield  {author} {\bibinfo {author} {\bibfnamefont {L.}~\bibnamefont
  {M{\"u}hlbacher}}\ and\ \bibinfo {author} {\bibfnamefont {E.}~\bibnamefont
  {Rabani}},\ }\href@noop {} {\bibfield  {journal} {\bibinfo  {journal}
  {Physical review letters}\ }\textbf {\bibinfo {volume} {100}},\ \bibinfo
  {pages} {176403} (\bibinfo {year} {2008})}\BibitemShut {NoStop}%
\bibitem [{\citenamefont {Weiss}\ \emph {et~al.}(2009)\citenamefont {Weiss},
  \citenamefont {Eckel}, \citenamefont {Thorwart},\ and\ \citenamefont
  {Egger}}]{Wei09}%
  \BibitemOpen
  \bibfield  {author} {\bibinfo {author} {\bibfnamefont {S.}~\bibnamefont
  {Weiss}}, \bibinfo {author} {\bibfnamefont {J.}~\bibnamefont {Eckel}},
  \bibinfo {author} {\bibfnamefont {M.}~\bibnamefont {Thorwart}}, \ and\
  \bibinfo {author} {\bibfnamefont {R.}~\bibnamefont {Egger}},\ }\href@noop {}
  {\bibfield  {journal} {\bibinfo  {journal} {Physical Review B}\ }\textbf
  {\bibinfo {volume} {79}},\ \bibinfo {pages} {249901} (\bibinfo {year}
  {2009})}\BibitemShut {NoStop}%
\bibitem [{\citenamefont {Segal}\ \emph {et~al.}(2010)\citenamefont {Segal},
  \citenamefont {Millis},\ and\ \citenamefont {Reichman}}]{Seg10b}%
  \BibitemOpen
  \bibfield  {author} {\bibinfo {author} {\bibfnamefont {D.}~\bibnamefont
  {Segal}}, \bibinfo {author} {\bibfnamefont {A.~J.}\ \bibnamefont {Millis}}, \
  and\ \bibinfo {author} {\bibfnamefont {D.~R.}\ \bibnamefont {Reichman}},\
  }\href@noop {} {\bibfield  {journal} {\bibinfo  {journal} {Physical Review
  B}\ }\textbf {\bibinfo {volume} {82}},\ \bibinfo {pages} {205323} (\bibinfo
  {year} {2010})}\BibitemShut {NoStop}%
\bibitem [{\citenamefont {Li}\ \emph {et~al.}(2020{\natexlab{a}})\citenamefont
  {Li}, \citenamefont {Govind}, \citenamefont {Isborn}, \citenamefont
  {DePrince},\ and\ \citenamefont {Lopata}}]{li_real-time_2020}%
  \BibitemOpen
  \bibfield  {author} {\bibinfo {author} {\bibfnamefont {X.}~\bibnamefont
  {Li}}, \bibinfo {author} {\bibfnamefont {N.}~\bibnamefont {Govind}}, \bibinfo
  {author} {\bibfnamefont {C.}~\bibnamefont {Isborn}}, \bibinfo {author}
  {\bibfnamefont {A.~E.}\ \bibnamefont {DePrince}}, \ and\ \bibinfo {author}
  {\bibfnamefont {K.}~\bibnamefont {Lopata}},\ }\href {\doibase
  10.1021/acs.chemrev.0c00223} {\bibfield  {journal} {\bibinfo  {journal}
  {Chemical Reviews}\ }\textbf {\bibinfo {volume} {120}},\ \bibinfo {pages}
  {9951} (\bibinfo {year} {2020}{\natexlab{a}})}\BibitemShut {NoStop}%
\bibitem [{\citenamefont {Goings}\ \emph {et~al.}(2018)\citenamefont {Goings},
  \citenamefont {Lestrange},\ and\ \citenamefont {Li}}]{goings_real-time_2018}%
  \BibitemOpen
  \bibfield  {author} {\bibinfo {author} {\bibfnamefont {J.~J.}\ \bibnamefont
  {Goings}}, \bibinfo {author} {\bibfnamefont {P.~J.}\ \bibnamefont
  {Lestrange}}, \ and\ \bibinfo {author} {\bibfnamefont {X.}~\bibnamefont
  {Li}},\ }\href {\doibase 10.1002/wcms.1341} {\bibfield  {journal} {\bibinfo
  {journal} {WIREs Computational Molecular Science}\ }\textbf {\bibinfo
  {volume} {8}},\ \bibinfo {pages} {e1341} (\bibinfo {year} {2018})},\ \bibinfo
  {note} {\_eprint:
  https://onlinelibrary.wiley.com/doi/pdf/10.1002/wcms.1341}\BibitemShut
  {NoStop}%
\bibitem [{\citenamefont {Schlünzen}\ \emph {et~al.}(2019)\citenamefont
  {Schlünzen}, \citenamefont {Hermanns}, \citenamefont {Scharnke},\ and\
  \citenamefont {Bonitz}}]{schlunzen_ultrafast_2019}%
  \BibitemOpen
  \bibfield  {author} {\bibinfo {author} {\bibfnamefont {N.}~\bibnamefont
  {Schlünzen}}, \bibinfo {author} {\bibfnamefont {S.}~\bibnamefont
  {Hermanns}}, \bibinfo {author} {\bibfnamefont {M.}~\bibnamefont {Scharnke}},
  \ and\ \bibinfo {author} {\bibfnamefont {M.}~\bibnamefont {Bonitz}},\ }\href
  {\doibase 10.1088/1361-648X/ab2d32} {\bibfield  {journal} {\bibinfo
  {journal} {Journal of Physics: Condensed Matter}\ }\textbf {\bibinfo {volume}
  {32}},\ \bibinfo {pages} {103001} (\bibinfo {year} {2019})}\BibitemShut
  {NoStop}%
\bibitem [{\citenamefont {Church}\ and\ \citenamefont
  {Rubenstein}(2021)}]{church_real-time_2021}%
  \BibitemOpen
  \bibfield  {author} {\bibinfo {author} {\bibfnamefont {M.~S.}\ \bibnamefont
  {Church}}\ and\ \bibinfo {author} {\bibfnamefont {B.~M.}\ \bibnamefont
  {Rubenstein}},\ }\href {\doibase 10.1063/5.0049116} {\bibfield  {journal}
  {\bibinfo  {journal} {The Journal of Chemical Physics}\ }\textbf {\bibinfo
  {volume} {154}},\ \bibinfo {pages} {184103} (\bibinfo {year}
  {2021})}\BibitemShut {NoStop}%
\bibitem [{\citenamefont {Thiel}(2014)}]{thiel_semiempirical_2014}%
  \BibitemOpen
  \bibfield  {author} {\bibinfo {author} {\bibfnamefont {W.}~\bibnamefont
  {Thiel}},\ }\href {\doibase 10.1002/wcms.1161} {\bibfield  {journal}
  {\bibinfo  {journal} {WIREs Computational Molecular Science}\ }\textbf
  {\bibinfo {volume} {4}},\ \bibinfo {pages} {145} (\bibinfo {year} {2014})},\
  \bibinfo {note} {\_eprint:
  https://onlinelibrary.wiley.com/doi/pdf/10.1002/wcms.1161}\BibitemShut
  {NoStop}%
\bibitem [{\citenamefont {Chatterjee}\ and\ \citenamefont
  {Makri}(2019)}]{chatterjee_real-time_2019}%
  \BibitemOpen
  \bibfield  {author} {\bibinfo {author} {\bibfnamefont {S.}~\bibnamefont
  {Chatterjee}}\ and\ \bibinfo {author} {\bibfnamefont {N.}~\bibnamefont
  {Makri}},\ }\href {\doibase 10.1021/acs.jpcb.9b08429} {\bibfield  {journal}
  {\bibinfo  {journal} {The Journal of Physical Chemistry B}\ }\textbf
  {\bibinfo {volume} {123}},\ \bibinfo {pages} {10470} (\bibinfo {year}
  {2019})}\BibitemShut {NoStop}%
\bibitem [{\citenamefont {Helgaker}\ \emph {et~al.}(2012)\citenamefont
  {Helgaker}, \citenamefont {Coriani}, \citenamefont {Jørgensen},
  \citenamefont {Kristensen}, \citenamefont {Olsen},\ and\ \citenamefont
  {Ruud}}]{helgaker_recent_2012}%
  \BibitemOpen
  \bibfield  {author} {\bibinfo {author} {\bibfnamefont {T.}~\bibnamefont
  {Helgaker}}, \bibinfo {author} {\bibfnamefont {S.}~\bibnamefont {Coriani}},
  \bibinfo {author} {\bibfnamefont {P.}~\bibnamefont {Jørgensen}}, \bibinfo
  {author} {\bibfnamefont {K.}~\bibnamefont {Kristensen}}, \bibinfo {author}
  {\bibfnamefont {J.}~\bibnamefont {Olsen}}, \ and\ \bibinfo {author}
  {\bibfnamefont {K.}~\bibnamefont {Ruud}},\ }\href {\doibase
  10.1021/cr2002239} {\bibfield  {journal} {\bibinfo  {journal} {Chemical
  Reviews}\ }\textbf {\bibinfo {volume} {112}},\ \bibinfo {pages} {543}
  (\bibinfo {year} {2012})}\BibitemShut {NoStop}%
\bibitem [{\citenamefont {Norman}\ and\ \citenamefont
  {Dreuw}(2018)}]{norman_simulating_2018}%
  \BibitemOpen
  \bibfield  {author} {\bibinfo {author} {\bibfnamefont {P.}~\bibnamefont
  {Norman}}\ and\ \bibinfo {author} {\bibfnamefont {A.}~\bibnamefont {Dreuw}},\
  }\href {\doibase 10.1021/acs.chemrev.8b00156} {\bibfield  {journal} {\bibinfo
   {journal} {Chemical Reviews}\ }\textbf {\bibinfo {volume} {118}},\ \bibinfo
  {pages} {7208} (\bibinfo {year} {2018})}\BibitemShut {NoStop}%
\bibitem [{\citenamefont {V.~Toukach}\ and\ \citenamefont
  {P.~Ananikov}(2013)}]{vtoukach_recent_2013}%
  \BibitemOpen
  \bibfield  {author} {\bibinfo {author} {\bibfnamefont {F.}~\bibnamefont
  {V.~Toukach}}\ and\ \bibinfo {author} {\bibfnamefont {V.}~\bibnamefont
  {P.~Ananikov}},\ }\href {\doibase 10.1039/C3CS60073D} {\bibfield  {journal}
  {\bibinfo  {journal} {Chemical Society Reviews}\ }\textbf {\bibinfo {volume}
  {42}},\ \bibinfo {pages} {8376} (\bibinfo {year} {2013})}\BibitemShut
  {NoStop}%
\bibitem [{\citenamefont {Baiardi}\ and\ \citenamefont {Reiher}(2019)}]{Bai19}%
  \BibitemOpen
  \bibfield  {author} {\bibinfo {author} {\bibfnamefont {A.}~\bibnamefont
  {Baiardi}}\ and\ \bibinfo {author} {\bibfnamefont {M.}~\bibnamefont
  {Reiher}},\ }\href@noop {} {\bibfield  {journal} {\bibinfo  {journal}
  {Journal of chemical theory and computation}\ }\textbf {\bibinfo {volume}
  {15}},\ \bibinfo {pages} {3481} (\bibinfo {year} {2019})}\BibitemShut
  {NoStop}%
\bibitem [{\citenamefont {Frahm}\ and\ \citenamefont
  {Pfannkuche}(2019)}]{Fra19}%
  \BibitemOpen
  \bibfield  {author} {\bibinfo {author} {\bibfnamefont {L.-H.}\ \bibnamefont
  {Frahm}}\ and\ \bibinfo {author} {\bibfnamefont {D.}~\bibnamefont
  {Pfannkuche}},\ }\href@noop {} {\bibfield  {journal} {\bibinfo  {journal}
  {Journal of Chemical Theory and Computation}\ }\textbf {\bibinfo {volume}
  {15}},\ \bibinfo {pages} {2154} (\bibinfo {year} {2019})}\BibitemShut
  {NoStop}%
\bibitem [{\citenamefont {Baiardi}(2021)}]{Bai21}%
  \BibitemOpen
  \bibfield  {author} {\bibinfo {author} {\bibfnamefont {A.}~\bibnamefont
  {Baiardi}},\ }\href@noop {} {\bibfield  {journal} {\bibinfo  {journal}
  {Journal of Chemical Theory and Computation}\ }\textbf {\bibinfo {volume}
  {17}},\ \bibinfo {pages} {3320} (\bibinfo {year} {2021})}\BibitemShut
  {NoStop}%
\bibitem [{\citenamefont {Li}\ \emph {et~al.}(2020{\natexlab{b}})\citenamefont
  {Li}, \citenamefont {Ren},\ and\ \citenamefont
  {Shuai}}]{li_finite-temperature_2020}%
  \BibitemOpen
  \bibfield  {author} {\bibinfo {author} {\bibfnamefont {W.}~\bibnamefont
  {Li}}, \bibinfo {author} {\bibfnamefont {J.}~\bibnamefont {Ren}}, \ and\
  \bibinfo {author} {\bibfnamefont {Z.}~\bibnamefont {Shuai}},\ }\href
  {\doibase 10.1021/acs.jpclett.0c01072} {\bibfield  {journal} {\bibinfo
  {journal} {The Journal of Physical Chemistry Letters}\ }\textbf {\bibinfo
  {volume} {11}},\ \bibinfo {pages} {4930} (\bibinfo {year}
  {2020}{\natexlab{b}})}\BibitemShut {NoStop}%
\bibitem [{\citenamefont {Sonk}\ \emph {et~al.}(2011)\citenamefont {Sonk},
  \citenamefont {Caricato},\ and\ \citenamefont {Schlegel}}]{sonk_td-ci_2011}%
  \BibitemOpen
  \bibfield  {author} {\bibinfo {author} {\bibfnamefont {J.~A.}\ \bibnamefont
  {Sonk}}, \bibinfo {author} {\bibfnamefont {M.}~\bibnamefont {Caricato}}, \
  and\ \bibinfo {author} {\bibfnamefont {H.~B.}\ \bibnamefont {Schlegel}},\
  }\href {\doibase 10.1021/jp107384p} {\bibfield  {journal} {\bibinfo
  {journal} {The Journal of Physical Chemistry A}\ }\textbf {\bibinfo {volume}
  {115}},\ \bibinfo {pages} {4678} (\bibinfo {year} {2011})}\BibitemShut
  {NoStop}%
\bibitem [{\citenamefont {Durden}\ and\ \citenamefont {Levine}(2022)}]{Dur22}%
  \BibitemOpen
  \bibfield  {author} {\bibinfo {author} {\bibfnamefont {A.~S.}\ \bibnamefont
  {Durden}}\ and\ \bibinfo {author} {\bibfnamefont {B.~G.}\ \bibnamefont
  {Levine}},\ }\href@noop {} {\bibfield  {journal} {\bibinfo  {journal}
  {Journal of Chemical Theory and Computation}\ }\textbf {\bibinfo {volume}
  {18}},\ \bibinfo {pages} {795} (\bibinfo {year} {2022})}\BibitemShut
  {NoStop}%
\bibitem [{\citenamefont {Peng}\ \emph {et~al.}(2018)\citenamefont {Peng},
  \citenamefont {Fales},\ and\ \citenamefont {Levine}}]{Pen18}%
  \BibitemOpen
  \bibfield  {author} {\bibinfo {author} {\bibfnamefont {W.-T.}\ \bibnamefont
  {Peng}}, \bibinfo {author} {\bibfnamefont {B.~S.}\ \bibnamefont {Fales}}, \
  and\ \bibinfo {author} {\bibfnamefont {B.~G.}\ \bibnamefont {Levine}},\
  }\href@noop {} {\bibfield  {journal} {\bibinfo  {journal} {Journal of
  chemical theory and computation}\ }\textbf {\bibinfo {volume} {14}},\
  \bibinfo {pages} {4129} (\bibinfo {year} {2018})}\BibitemShut {NoStop}%
\bibitem [{\citenamefont {Sato}\ and\ \citenamefont
  {Ishikawa}(2013{\natexlab{a}})}]{Sat13}%
  \BibitemOpen
  \bibfield  {author} {\bibinfo {author} {\bibfnamefont {T.}~\bibnamefont
  {Sato}}\ and\ \bibinfo {author} {\bibfnamefont {K.~L.}\ \bibnamefont
  {Ishikawa}},\ }\href@noop {} {\bibfield  {journal} {\bibinfo  {journal}
  {Phys. Rev. A}\ }\textbf {\bibinfo {volume} {88}},\ \bibinfo {pages} {023402}
  (\bibinfo {year} {2013}{\natexlab{a}})}\BibitemShut {NoStop}%
\bibitem [{\citenamefont {Miranda}\ \emph {et~al.}(2011)\citenamefont
  {Miranda}, \citenamefont {Fisher}, \citenamefont {Stella},\ and\
  \citenamefont {Horsfield}}]{mir11}%
  \BibitemOpen
  \bibfield  {author} {\bibinfo {author} {\bibfnamefont {R.~P.}\ \bibnamefont
  {Miranda}}, \bibinfo {author} {\bibfnamefont {A.~J.}\ \bibnamefont {Fisher}},
  \bibinfo {author} {\bibfnamefont {L.}~\bibnamefont {Stella}}, \ and\ \bibinfo
  {author} {\bibfnamefont {A.~P.}\ \bibnamefont {Horsfield}},\ }\href@noop {}
  {\bibfield  {journal} {\bibinfo  {journal} {J. Chem. Phys.}\ }\textbf
  {\bibinfo {volume} {134}},\ \bibinfo {pages} {244101} (\bibinfo {year}
  {2011})}\BibitemShut {NoStop}%
\bibitem [{\citenamefont {Kochetov}\ and\ \citenamefont
  {Bokarev}(2022)}]{kochetov_rhodyn_2022}%
  \BibitemOpen
  \bibfield  {author} {\bibinfo {author} {\bibfnamefont {V.}~\bibnamefont
  {Kochetov}}\ and\ \bibinfo {author} {\bibfnamefont {S.~I.}\ \bibnamefont
  {Bokarev}},\ }\href {\doibase 10.1021/acs.jctc.1c01097} {\bibfield  {journal}
  {\bibinfo  {journal} {Journal of Chemical Theory and Computation}\ }\textbf
  {\bibinfo {volume} {18}},\ \bibinfo {pages} {46} (\bibinfo {year}
  {2022})}\BibitemShut {NoStop}%
\bibitem [{\citenamefont {Isborn}\ \emph {et~al.}(2007)\citenamefont {Isborn},
  \citenamefont {Li},\ and\ \citenamefont
  {Tully}}]{isborn_time-dependent_2007-1}%
  \BibitemOpen
  \bibfield  {author} {\bibinfo {author} {\bibfnamefont {C.~M.}\ \bibnamefont
  {Isborn}}, \bibinfo {author} {\bibfnamefont {X.}~\bibnamefont {Li}}, \ and\
  \bibinfo {author} {\bibfnamefont {J.~C.}\ \bibnamefont {Tully}},\ }\href
  {\doibase 10.1063/1.2713391} {\bibfield  {journal} {\bibinfo  {journal} {The
  Journal of Chemical Physics}\ }\textbf {\bibinfo {volume} {126}},\ \bibinfo
  {pages} {134307} (\bibinfo {year} {2007})}\BibitemShut {NoStop}%
\bibitem [{\citenamefont {Knizia}\ and\ \citenamefont {Chan}(2012)}]{Kni12}%
  \BibitemOpen
  \bibfield  {author} {\bibinfo {author} {\bibfnamefont {G.}~\bibnamefont
  {Knizia}}\ and\ \bibinfo {author} {\bibfnamefont {G.~K.-L.}\ \bibnamefont
  {Chan}},\ }\href@noop {} {\bibfield  {journal} {\bibinfo  {journal} {Phys.
  Rev. Lett.}\ }\textbf {\bibinfo {volume} {109}},\ \bibinfo {pages} {186404}
  (\bibinfo {year} {2012})}\BibitemShut {NoStop}%
\bibitem [{\citenamefont {Knizia}\ and\ \citenamefont {Chan}(2013)}]{Kni13}%
  \BibitemOpen
  \bibfield  {author} {\bibinfo {author} {\bibfnamefont {G.}~\bibnamefont
  {Knizia}}\ and\ \bibinfo {author} {\bibfnamefont {G.~K.-L.}\ \bibnamefont
  {Chan}},\ }\href@noop {} {\bibfield  {journal} {\bibinfo  {journal} {J. Chem.
  Theory and Comput.}\ }\textbf {\bibinfo {volume} {9}},\ \bibinfo {pages}
  {1428} (\bibinfo {year} {2013})}\BibitemShut {NoStop}%
\bibitem [{\citenamefont {Wouters}\ \emph
  {et~al.}(2016{\natexlab{a}})\citenamefont {Wouters}, \citenamefont
  {Jim\'{e}nez-Hoyos}, \citenamefont {Sun},\ and\ \citenamefont
  {Chan}}]{Wou16}%
  \BibitemOpen
  \bibfield  {author} {\bibinfo {author} {\bibfnamefont {S.}~\bibnamefont
  {Wouters}}, \bibinfo {author} {\bibfnamefont {C.~A.}\ \bibnamefont
  {Jim\'{e}nez-Hoyos}}, \bibinfo {author} {\bibfnamefont {Q.}~\bibnamefont
  {Sun}}, \ and\ \bibinfo {author} {\bibfnamefont {G.~K.-L.}\ \bibnamefont
  {Chan}},\ }\href@noop {} {\bibfield  {journal} {\bibinfo  {journal} {J. Chem.
  Theory and Comput.}\ }\textbf {\bibinfo {volume} {12}},\ \bibinfo {pages}
  {2706} (\bibinfo {year} {2016}{\natexlab{a}})}\BibitemShut {NoStop}%
\bibitem [{\citenamefont {Wu}\ \emph {et~al.}(2019)\citenamefont {Wu},
  \citenamefont {Cui}, \citenamefont {Tong}, \citenamefont {Lindsey},
  \citenamefont {Chan},\ and\ \citenamefont {Lin}}]{wu_projected_2019}%
  \BibitemOpen
  \bibfield  {author} {\bibinfo {author} {\bibfnamefont {X.}~\bibnamefont
  {Wu}}, \bibinfo {author} {\bibfnamefont {Z.-H.}\ \bibnamefont {Cui}},
  \bibinfo {author} {\bibfnamefont {Y.}~\bibnamefont {Tong}}, \bibinfo {author}
  {\bibfnamefont {M.}~\bibnamefont {Lindsey}}, \bibinfo {author} {\bibfnamefont
  {G.~K.-L.}\ \bibnamefont {Chan}}, \ and\ \bibinfo {author} {\bibfnamefont
  {L.}~\bibnamefont {Lin}},\ }\href {\doibase 10.1063/1.5108818} {\bibfield
  {journal} {\bibinfo  {journal} {The Journal of Chemical Physics}\ }\textbf
  {\bibinfo {volume} {151}},\ \bibinfo {pages} {064108} (\bibinfo {year}
  {2019})}\BibitemShut {NoStop}%
\bibitem [{\citenamefont {Zheng}\ and\ \citenamefont
  {Chan}(2016)}]{zheng_ground-state_2016}%
  \BibitemOpen
  \bibfield  {author} {\bibinfo {author} {\bibfnamefont {B.-X.}\ \bibnamefont
  {Zheng}}\ and\ \bibinfo {author} {\bibfnamefont {G.~K.-L.}\ \bibnamefont
  {Chan}},\ }\href {\doibase 10.1103/PhysRevB.93.035126} {\bibfield  {journal}
  {\bibinfo  {journal} {Physical Review B}\ }\textbf {\bibinfo {volume} {93}},\
  \bibinfo {pages} {035126} (\bibinfo {year} {2016})}\BibitemShut {NoStop}%
\bibitem [{\citenamefont {Cui}\ \emph {et~al.}(2020)\citenamefont {Cui},
  \citenamefont {Zhu},\ and\ \citenamefont {Chan}}]{cui_efficient_2020}%
  \BibitemOpen
  \bibfield  {author} {\bibinfo {author} {\bibfnamefont {Z.-H.}\ \bibnamefont
  {Cui}}, \bibinfo {author} {\bibfnamefont {T.}~\bibnamefont {Zhu}}, \ and\
  \bibinfo {author} {\bibfnamefont {G.~K.-L.}\ \bibnamefont {Chan}},\ }\href
  {\doibase 10.1021/acs.jctc.9b00933} {\bibfield  {journal} {\bibinfo
  {journal} {Journal of Chemical Theory and Computation}\ }\textbf {\bibinfo
  {volume} {16}},\ \bibinfo {pages} {119} (\bibinfo {year} {2020})}\BibitemShut
  {NoStop}%
\bibitem [{\citenamefont {Faulstich}\ \emph {et~al.}(2022)\citenamefont
  {Faulstich}, \citenamefont {Kim}, \citenamefont {Cui}, \citenamefont {Wen},
  \citenamefont {Kin-Lic~Chan},\ and\ \citenamefont
  {Lin}}]{faulstich_pure_2022}%
  \BibitemOpen
  \bibfield  {author} {\bibinfo {author} {\bibfnamefont {F.~M.}\ \bibnamefont
  {Faulstich}}, \bibinfo {author} {\bibfnamefont {R.}~\bibnamefont {Kim}},
  \bibinfo {author} {\bibfnamefont {Z.-H.}\ \bibnamefont {Cui}}, \bibinfo
  {author} {\bibfnamefont {Z.}~\bibnamefont {Wen}}, \bibinfo {author}
  {\bibfnamefont {G.}~\bibnamefont {Kin-Lic~Chan}}, \ and\ \bibinfo {author}
  {\bibfnamefont {L.}~\bibnamefont {Lin}},\ }\href {\doibase
  10.1021/acs.jctc.1c01061} {\bibfield  {journal} {\bibinfo  {journal} {Journal
  of Chemical Theory and Computation}\ }\textbf {\bibinfo {volume} {18}},\
  \bibinfo {pages} {851} (\bibinfo {year} {2022})}\BibitemShut {NoStop}%
\bibitem [{\citenamefont {Reinhard}\ \emph {et~al.}(2019)\citenamefont
  {Reinhard}, \citenamefont {Mordovina}, \citenamefont {Hubig}, \citenamefont
  {Kretchmer}, \citenamefont {Schollwöck}, \citenamefont {Appel},
  \citenamefont {Sentef},\ and\ \citenamefont
  {Rubio}}]{reinhard_density-matrix_2019}%
  \BibitemOpen
  \bibfield  {author} {\bibinfo {author} {\bibfnamefont {T.~E.}\ \bibnamefont
  {Reinhard}}, \bibinfo {author} {\bibfnamefont {U.}~\bibnamefont {Mordovina}},
  \bibinfo {author} {\bibfnamefont {C.}~\bibnamefont {Hubig}}, \bibinfo
  {author} {\bibfnamefont {J.~S.}\ \bibnamefont {Kretchmer}}, \bibinfo {author}
  {\bibfnamefont {U.}~\bibnamefont {Schollwöck}}, \bibinfo {author}
  {\bibfnamefont {H.}~\bibnamefont {Appel}}, \bibinfo {author} {\bibfnamefont
  {M.~A.}\ \bibnamefont {Sentef}}, \ and\ \bibinfo {author} {\bibfnamefont
  {A.}~\bibnamefont {Rubio}},\ }\href {\doibase 10.1021/acs.jctc.8b01116}
  {\bibfield  {journal} {\bibinfo  {journal} {Journal of Chemical Theory and
  Computation}\ }\textbf {\bibinfo {volume} {15}},\ \bibinfo {pages} {2221}
  (\bibinfo {year} {2019})}\BibitemShut {NoStop}%
\bibitem [{\citenamefont {Pham}\ \emph {et~al.}(2020)\citenamefont {Pham},
  \citenamefont {Hermes},\ and\ \citenamefont
  {Gagliardi}}]{pham_periodic_2020}%
  \BibitemOpen
  \bibfield  {author} {\bibinfo {author} {\bibfnamefont {H.~Q.}\ \bibnamefont
  {Pham}}, \bibinfo {author} {\bibfnamefont {M.~R.}\ \bibnamefont {Hermes}}, \
  and\ \bibinfo {author} {\bibfnamefont {L.}~\bibnamefont {Gagliardi}},\ }\href
  {\doibase 10.1021/acs.jctc.9b00939} {\bibfield  {journal} {\bibinfo
  {journal} {Journal of Chemical Theory and Computation}\ }\textbf {\bibinfo
  {volume} {16}},\ \bibinfo {pages} {130} (\bibinfo {year} {2020})}\BibitemShut
  {NoStop}%
\bibitem [{\citenamefont {Fan}\ and\ \citenamefont
  {Jie}(2015)}]{fan_cluster_2015}%
  \BibitemOpen
  \bibfield  {author} {\bibinfo {author} {\bibfnamefont {Z.}~\bibnamefont
  {Fan}}\ and\ \bibinfo {author} {\bibfnamefont {Q.-l.}\ \bibnamefont {Jie}},\
  }\href {\doibase 10.1103/PhysRevB.91.195118} {\bibfield  {journal} {\bibinfo
  {journal} {Physical Review B}\ }\textbf {\bibinfo {volume} {91}},\ \bibinfo
  {pages} {195118} (\bibinfo {year} {2015})}\BibitemShut {NoStop}%
\bibitem [{\citenamefont {Gunst}\ \emph {et~al.}(2017)\citenamefont {Gunst},
  \citenamefont {Wouters}, \citenamefont {De~Baerdemacker},\ and\ \citenamefont
  {Van~Neck}}]{gunst_block_2017}%
  \BibitemOpen
  \bibfield  {author} {\bibinfo {author} {\bibfnamefont {K.}~\bibnamefont
  {Gunst}}, \bibinfo {author} {\bibfnamefont {S.}~\bibnamefont {Wouters}},
  \bibinfo {author} {\bibfnamefont {S.}~\bibnamefont {De~Baerdemacker}}, \ and\
  \bibinfo {author} {\bibfnamefont {D.}~\bibnamefont {Van~Neck}},\ }\href
  {\doibase 10.1103/PhysRevB.95.195127} {\bibfield  {journal} {\bibinfo
  {journal} {Physical Review B}\ }\textbf {\bibinfo {volume} {95}},\ \bibinfo
  {pages} {195127} (\bibinfo {year} {2017})}\BibitemShut {NoStop}%
\bibitem [{\citenamefont {Tran}\ \emph {et~al.}(2019)\citenamefont {Tran},
  \citenamefont {Voorhis},\ and\ \citenamefont {Thom}}]{Tra19}%
  \BibitemOpen
  \bibfield  {author} {\bibinfo {author} {\bibfnamefont {H.~K.}\ \bibnamefont
  {Tran}}, \bibinfo {author} {\bibfnamefont {T.~V.}\ \bibnamefont {Voorhis}}, \
  and\ \bibinfo {author} {\bibfnamefont {A.~J.~W.}\ \bibnamefont {Thom}},\
  }\href@noop {} {\bibfield  {journal} {\bibinfo  {journal} {J. Chem. Phys.}\
  }\textbf {\bibinfo {volume} {151}},\ \bibinfo {pages} {034112} (\bibinfo
  {year} {2019})}\BibitemShut {NoStop}%
\bibitem [{\citenamefont {Mitra}\ \emph {et~al.}(2021)\citenamefont {Mitra},
  \citenamefont {Pham}, \citenamefont {Pandharkar}, \citenamefont {Hermes},\
  and\ \citenamefont {Gagliardi}}]{mitra_excited_2021}%
  \BibitemOpen
  \bibfield  {author} {\bibinfo {author} {\bibfnamefont {A.}~\bibnamefont
  {Mitra}}, \bibinfo {author} {\bibfnamefont {H.~Q.}\ \bibnamefont {Pham}},
  \bibinfo {author} {\bibfnamefont {R.}~\bibnamefont {Pandharkar}}, \bibinfo
  {author} {\bibfnamefont {M.~R.}\ \bibnamefont {Hermes}}, \ and\ \bibinfo
  {author} {\bibfnamefont {L.}~\bibnamefont {Gagliardi}},\ }\href {\doibase
  10.1021/acs.jpclett.1c03229} {\bibfield  {journal} {\bibinfo  {journal} {The
  Journal of Physical Chemistry Letters}\ ,\ \bibinfo {pages} {11688}}
  (\bibinfo {year} {2021})}\BibitemShut {NoStop}%
\bibitem [{\citenamefont {Qiao}\ and\ \citenamefont
  {Jie}(2021)}]{qiao_density_2021}%
  \BibitemOpen
  \bibfield  {author} {\bibinfo {author} {\bibfnamefont {J.}~\bibnamefont
  {Qiao}}\ and\ \bibinfo {author} {\bibfnamefont {Q.}~\bibnamefont {Jie}},\
  }\href {\doibase 10.1016/j.cpc.2020.107712} {\bibfield  {journal} {\bibinfo
  {journal} {Computer Physics Communications}\ }\textbf {\bibinfo {volume}
  {261}},\ \bibinfo {pages} {107712} (\bibinfo {year} {2021})}\BibitemShut
  {NoStop}%
\bibitem [{\citenamefont {Pham}\ \emph {et~al.}(2018)\citenamefont {Pham},
  \citenamefont {Bernales},\ and\ \citenamefont {Gagliardi}}]{pham_can_2018}%
  \BibitemOpen
  \bibfield  {author} {\bibinfo {author} {\bibfnamefont {H.~Q.}\ \bibnamefont
  {Pham}}, \bibinfo {author} {\bibfnamefont {V.}~\bibnamefont {Bernales}}, \
  and\ \bibinfo {author} {\bibfnamefont {L.}~\bibnamefont {Gagliardi}},\ }\href
  {\doibase 10.1021/acs.jctc.7b01248} {\bibfield  {journal} {\bibinfo
  {journal} {Journal of Chemical Theory and Computation}\ }\textbf {\bibinfo
  {volume} {14}},\ \bibinfo {pages} {1960} (\bibinfo {year}
  {2018})}\BibitemShut {NoStop}%
\bibitem [{\citenamefont {Sandhoefer}\ and\ \citenamefont
  {Chan}(2016)}]{sandhoefer_density_2016}%
  \BibitemOpen
  \bibfield  {author} {\bibinfo {author} {\bibfnamefont {B.}~\bibnamefont
  {Sandhoefer}}\ and\ \bibinfo {author} {\bibfnamefont {G.~K.-L.}\ \bibnamefont
  {Chan}},\ }\href {\doibase 10.1103/PhysRevB.94.085115} {\bibfield  {journal}
  {\bibinfo  {journal} {Physical Review B}\ }\textbf {\bibinfo {volume} {94}},\
  \bibinfo {pages} {085115} (\bibinfo {year} {2016})}\BibitemShut {NoStop}%
\bibitem [{\citenamefont {Ricke}\ \emph {et~al.}(2017)\citenamefont {Ricke},
  \citenamefont {Welborn}, \citenamefont {Ye},\ and\ \citenamefont
  {Van~Voorhis}}]{ricke_performance_2017}%
  \BibitemOpen
  \bibfield  {author} {\bibinfo {author} {\bibfnamefont {N.}~\bibnamefont
  {Ricke}}, \bibinfo {author} {\bibfnamefont {M.}~\bibnamefont {Welborn}},
  \bibinfo {author} {\bibfnamefont {H.-Z.}\ \bibnamefont {Ye}}, \ and\ \bibinfo
  {author} {\bibfnamefont {T.}~\bibnamefont {Van~Voorhis}},\ }\href {\doibase
  10.1080/00268976.2017.1290839} {\bibfield  {journal} {\bibinfo  {journal}
  {Molecular Physics}\ }\textbf {\bibinfo {volume} {115}},\ \bibinfo {pages}
  {2242} (\bibinfo {year} {2017})}\BibitemShut {NoStop}%
\bibitem [{\citenamefont {Kretchmer}\ and\ \citenamefont
  {Chan}(2018{\natexlab{a}})}]{kretchmer_real-time_2018}%
  \BibitemOpen
  \bibfield  {author} {\bibinfo {author} {\bibfnamefont {J.~S.}\ \bibnamefont
  {Kretchmer}}\ and\ \bibinfo {author} {\bibfnamefont {G.~K.-L.}\ \bibnamefont
  {Chan}},\ }\href {\doibase 10.1063/1.5012766} {\bibfield  {journal} {\bibinfo
   {journal} {The Journal of Chemical Physics}\ }\textbf {\bibinfo {volume}
  {148}},\ \bibinfo {pages} {054108} (\bibinfo {year}
  {2018}{\natexlab{a}})}\BibitemShut {NoStop}%
\bibitem [{\citenamefont {Wouters}\ \emph
  {et~al.}(2016{\natexlab{b}})\citenamefont {Wouters}, \citenamefont
  {Jiménez-Hoyos}, \citenamefont {Sun},\ and\ \citenamefont
  {Chan}}]{wouters_practical_2016}%
  \BibitemOpen
  \bibfield  {author} {\bibinfo {author} {\bibfnamefont {S.}~\bibnamefont
  {Wouters}}, \bibinfo {author} {\bibfnamefont {C.~A.}\ \bibnamefont
  {Jiménez-Hoyos}}, \bibinfo {author} {\bibfnamefont {Q.}~\bibnamefont {Sun}},
  \ and\ \bibinfo {author} {\bibfnamefont {G.~K.-L.}\ \bibnamefont {Chan}},\
  }\href {\doibase 10.1021/acs.jctc.6b00316} {\bibfield  {journal} {\bibinfo
  {journal} {Journal of Chemical Theory and Computation}\ }\textbf {\bibinfo
  {volume} {12}},\ \bibinfo {pages} {2706} (\bibinfo {year}
  {2016}{\natexlab{b}})}\BibitemShut {NoStop}%
\bibitem [{\citenamefont {Sato}\ and\ \citenamefont
  {Ishikawa}(2013{\natexlab{b}})}]{sato_time-dependent_2013}%
  \BibitemOpen
  \bibfield  {author} {\bibinfo {author} {\bibfnamefont {T.}~\bibnamefont
  {Sato}}\ and\ \bibinfo {author} {\bibfnamefont {K.~L.}\ \bibnamefont
  {Ishikawa}},\ }\href {\doibase 10.1103/PhysRevA.88.023402} {\bibfield
  {journal} {\bibinfo  {journal} {Physical Review A}\ }\textbf {\bibinfo
  {volume} {88}},\ \bibinfo {pages} {023402} (\bibinfo {year}
  {2013}{\natexlab{b}})}\BibitemShut {NoStop}%
\bibitem [{\citenamefont {Kretchmer}\ and\ \citenamefont
  {Chan}(2018{\natexlab{b}})}]{Kre18b}%
  \BibitemOpen
  \bibfield  {author} {\bibinfo {author} {\bibfnamefont {J.~S.}\ \bibnamefont
  {Kretchmer}}\ and\ \bibinfo {author} {\bibfnamefont {G.~K.-L.}\ \bibnamefont
  {Chan}},\ }\href@noop {} {\bibfield  {journal} {\bibinfo  {journal} {J. Phys.
  Chem. Lett.}\ }\textbf {\bibinfo {volume} {9}},\ \bibinfo {pages} {2863}
  (\bibinfo {year} {2018}{\natexlab{b}})}\BibitemShut {NoStop}%
\bibitem [{\citenamefont {Manthe}\ \emph {et~al.}(1992)\citenamefont {Manthe},
  \citenamefont {Meyer},\ and\ \citenamefont
  {Cederbaum}}]{manthe_wavepacket_1992}%
  \BibitemOpen
  \bibfield  {author} {\bibinfo {author} {\bibfnamefont {U.}~\bibnamefont
  {Manthe}}, \bibinfo {author} {\bibfnamefont {H.}~\bibnamefont {Meyer}}, \
  and\ \bibinfo {author} {\bibfnamefont {L.~S.}\ \bibnamefont {Cederbaum}},\
  }\href {\doibase 10.1063/1.463007} {\bibfield  {journal} {\bibinfo  {journal}
  {The Journal of Chemical Physics}\ }\textbf {\bibinfo {volume} {97}},\
  \bibinfo {pages} {3199} (\bibinfo {year} {1992})}\BibitemShut {NoStop}%
\bibitem [{\citenamefont {Meyer}\ \emph {et~al.}(1990)\citenamefont {Meyer},
  \citenamefont {Manthe},\ and\ \citenamefont
  {Cederbaum}}]{meyer_multi-configurational_1990}%
  \BibitemOpen
  \bibfield  {author} {\bibinfo {author} {\bibfnamefont {H.~D.}\ \bibnamefont
  {Meyer}}, \bibinfo {author} {\bibfnamefont {U.}~\bibnamefont {Manthe}}, \
  and\ \bibinfo {author} {\bibfnamefont {L.~S.}\ \bibnamefont {Cederbaum}},\
  }\href {\doibase 10.1016/0009-2614(90)87014-I} {\bibfield  {journal}
  {\bibinfo  {journal} {Chemical Physics Letters}\ }\textbf {\bibinfo {volume}
  {165}},\ \bibinfo {pages} {73} (\bibinfo {year} {1990})}\BibitemShut
  {NoStop}%
\bibitem [{\citenamefont {Helgaker}\ \emph {et~al.}(2014)\citenamefont
  {Helgaker}, \citenamefont {Jorgensen},\ and\ \citenamefont {Olsen}}]{Hel14}%
  \BibitemOpen
  \bibfield  {author} {\bibinfo {author} {\bibfnamefont {T.}~\bibnamefont
  {Helgaker}}, \bibinfo {author} {\bibfnamefont {P.}~\bibnamefont {Jorgensen}},
  \ and\ \bibinfo {author} {\bibfnamefont {J.}~\bibnamefont {Olsen}},\
  }\href@noop {} {\emph {\bibinfo {title} {Molecular electronic-structure
  theory}}}\ (\bibinfo  {publisher} {John Wiley \& Sons},\ \bibinfo {year}
  {2014})\BibitemShut {NoStop}%
\bibitem [{\citenamefont {Sun}\ \emph {et~al.}(2020)\citenamefont {Sun},
  \citenamefont {Zhang}, \citenamefont {Banerjee}, \citenamefont {Bao},
  \citenamefont {Barbry}, \citenamefont {Blunt}, \citenamefont {Bogdanov},
  \citenamefont {Booth}, \citenamefont {Chen}, \citenamefont {Cui} \emph
  {et~al.}}]{sun2020recent}%
  \BibitemOpen
  \bibfield  {author} {\bibinfo {author} {\bibfnamefont {Q.}~\bibnamefont
  {Sun}}, \bibinfo {author} {\bibfnamefont {X.}~\bibnamefont {Zhang}}, \bibinfo
  {author} {\bibfnamefont {S.}~\bibnamefont {Banerjee}}, \bibinfo {author}
  {\bibfnamefont {P.}~\bibnamefont {Bao}}, \bibinfo {author} {\bibfnamefont
  {M.}~\bibnamefont {Barbry}}, \bibinfo {author} {\bibfnamefont {N.~S.}\
  \bibnamefont {Blunt}}, \bibinfo {author} {\bibfnamefont {N.~A.}\ \bibnamefont
  {Bogdanov}}, \bibinfo {author} {\bibfnamefont {G.~H.}\ \bibnamefont {Booth}},
  \bibinfo {author} {\bibfnamefont {J.}~\bibnamefont {Chen}}, \bibinfo {author}
  {\bibfnamefont {Z.-H.}\ \bibnamefont {Cui}},  \emph {et~al.},\ }\href@noop {}
  {\bibfield  {journal} {\bibinfo  {journal} {The Journal of chemical physics}\
  }\textbf {\bibinfo {volume} {153}},\ \bibinfo {pages} {024109} (\bibinfo
  {year} {2020})}\BibitemShut {NoStop}%
\bibitem [{rtd()}]{rtdmetGit_link}%
  \BibitemOpen
  \href@noop {} {}\bibinfo {howpublished}
  {\url{https://github.com/jskretchmer/real_time_pDMET}}\BibitemShut {NoStop}%
\bibitem [{\citenamefont {Titvinidze}\ \emph {et~al.}(2015)\citenamefont
  {Titvinidze}, \citenamefont {Schwabe},\ and\ \citenamefont
  {Potthoff}}]{titvinidze_strong-coupling_2015}%
  \BibitemOpen
  \bibfield  {author} {\bibinfo {author} {\bibfnamefont {I.}~\bibnamefont
  {Titvinidze}}, \bibinfo {author} {\bibfnamefont {A.}~\bibnamefont {Schwabe}},
  \ and\ \bibinfo {author} {\bibfnamefont {M.}~\bibnamefont {Potthoff}},\
  }\href {\doibase 10.1140/epjb/e2014-50772-1} {\bibfield  {journal} {\bibinfo
  {journal} {The European Physical Journal B}\ }\textbf {\bibinfo {volume}
  {88}},\ \bibinfo {pages} {9} (\bibinfo {year} {2015})}\BibitemShut {NoStop}%
\bibitem [{\citenamefont {Eickhoff}\ and\ \citenamefont
  {Anders}(2020)}]{eickhoff_strongly_2020}%
  \BibitemOpen
  \bibfield  {author} {\bibinfo {author} {\bibfnamefont {F.}~\bibnamefont
  {Eickhoff}}\ and\ \bibinfo {author} {\bibfnamefont {F.~B.}\ \bibnamefont
  {Anders}},\ }\href {\doibase 10.1103/PhysRevB.102.205132} {\bibfield
  {journal} {\bibinfo  {journal} {Physical Review B}\ }\textbf {\bibinfo
  {volume} {102}},\ \bibinfo {pages} {205132} (\bibinfo {year} {2020})},\
  \bibinfo {note} {arXiv:2008.09034 [cond-mat]}\BibitemShut {NoStop}%
\bibitem [{\citenamefont {Baiardi}\ and\ \citenamefont {Reiher}(2020)}]{Bai20}%
  \BibitemOpen
  \bibfield  {author} {\bibinfo {author} {\bibfnamefont {A.}~\bibnamefont
  {Baiardi}}\ and\ \bibinfo {author} {\bibfnamefont {M.}~\bibnamefont
  {Reiher}},\ }\href@noop {} {\bibfield  {journal} {\bibinfo  {journal} {The
  Journal of Chemical Physics}\ }\textbf {\bibinfo {volume} {152}},\ \bibinfo
  {pages} {040903} (\bibinfo {year} {2020})}\BibitemShut {NoStop}%
\bibitem [{\citenamefont {Shao}\ \emph {et~al.}(2020)\citenamefont {Shao},
  \citenamefont {Tohyama}, \citenamefont {Luo},\ and\ \citenamefont
  {Lu}}]{shao_2020}%
  \BibitemOpen
  \bibfield  {author} {\bibinfo {author} {\bibfnamefont {C.}~\bibnamefont
  {Shao}}, \bibinfo {author} {\bibfnamefont {T.}~\bibnamefont {Tohyama}},
  \bibinfo {author} {\bibfnamefont {H.-G.}\ \bibnamefont {Luo}}, \ and\
  \bibinfo {author} {\bibfnamefont {H.}~\bibnamefont {Lu}},\ }\href@noop {}
  {\bibfield  {journal} {\bibinfo  {journal} {Physical Review B}\ }\textbf
  {\bibinfo {volume} {101}},\ \bibinfo {pages} {045128} (\bibinfo {year}
  {2020})}\BibitemShut {NoStop}%
\bibitem [{\citenamefont {Fishman}\ \emph {et~al.}(2020)\citenamefont
  {Fishman}, \citenamefont {White},\ and\ \citenamefont
  {Stoudenmire}}]{itensor}%
  \BibitemOpen
  \bibfield  {author} {\bibinfo {author} {\bibfnamefont {M.}~\bibnamefont
  {Fishman}}, \bibinfo {author} {\bibfnamefont {S.~R.}\ \bibnamefont {White}},
  \ and\ \bibinfo {author} {\bibfnamefont {E.~M.}\ \bibnamefont
  {Stoudenmire}},\ }\href@noop {} {\enquote {\bibinfo {title} {The
  \mbox{ITensor} software library for tensor network calculations},}\ }
  (\bibinfo {year} {2020}),\ \Eprint {http://arxiv.org/abs/2007.14822}
  {arXiv:2007.14822} \BibitemShut {NoStop}%
\end{thebibliography}%
